\documentclass[%
reprint,
superscriptaddress,
amsmath,amssymb,
aps,
]{revtex4-2}
\usepackage{graphicx}
\usepackage{dcolumn}
\usepackage{bm}
\usepackage{xcolor}
\usepackage{siunitx}

\begin{document}

\title{Transparency, Nonclassicality and Nonreciprocity in Chiral Waveguide Quantum Electrodynamics}
\author{Qingtian Miao}
\email{qm8@tamu.edu}
\affiliation{Department of Physics and Astronomy, Texas A\&M University, Texas 77843, USA}
\author{G. S. Agarwal}
\email{Girish.Agarwal@ag.tamu.edu}
\affiliation{Institute for Quantum Science and Engineering, Department of Physics and Astronomy, Department of Biological and Agricultural Engineering, Texas A\&M University, College Station, Texas 77843, USA}

\date{\today}

\begin{abstract}
We examine quantum statistical properties of transmission and reflection from a chiral waveguide coupled to qubits for arbitrary input powers. We report on several remarkable features of output fields such as transparency, quantum nonreciprocity and the second-order correlation function $g^{(2)}(0)$ values less than unity. In particular, for two qubits detuned antisymmetrically with respect to the central waveguide frequency, we find transparency in forward transmission and in photon numbers for arbitrary values of the input powers provided the phase separation between qubits is an integer multiple of $\pi$. Values of $g^{(2)}(0)$ less than unity can be reached even for nonzero value of the intrinsic damping by using phase separation different from integer multiple of $\pi$, marking the transition from classical to quantum light. We also uncover a new type of quantum criticality that enables complete suppression of forward-propagating amplitude transmission at specific driving powers, giving rise to enhanced nonreciprocal effects in both transmission and quantum fluctuations in amplitudes. Forward propagation amplifies the quantum fluctuations in amplitudes, while backward propagation significantly suppresses them. These findings open new pathways for controlling light-matter interactions in chiral quantum electrodynamics, with potential applications in quantum information and nonreciprocal quantum devices.
\end{abstract}

\maketitle

\section{Introduction}
In recent years, chiral quantum optics has emerged as an important branch of quantum optics, garnering significant interest due to its inherently asymmetric light-matter interactions \cite{lodahl2017chiral,bliokh2012transverse,bliokh2015spin,aiello2015transverse,bliokh2015quantum}. These interactions leverage the spin-momentum locking of light, enabling a strong directional dependence in photon emission, scattering, and absorption processes. Experimental observations of directional emission have been made spanning various photonic nanostructures, ranging from dielectric to plasmonic platforms \cite{luxmoore2013interfacing,junge2013strong,luxmoore2013optical,neugebauer2014polarization,petersen2014chiral,rodriguez2014resolving,shomroni2014all,sollner2015deterministic,le2015nanophotonic,coles2016chirality,lee2012role,lin2013polarization,rodriguez2013near,o2014spin}. Complementing these developments, waveguide quantum electrodynamics (QED) has established itself as a versatile framework for studying light-matter interactions \cite{roy2017colloquium,sheremet2023waveguide}. Unlike free-space systems, waveguide QED provide structured environments that enable strong and tunable couplings between photons and quantum emitters, which is particularly promising for quantum information processing and entanglement generation \cite{kannan2020generating,gonzalez2014generation,facchi2016bound,shah2024stabilizing}. A shared dissipation channel enables long-range, photon-mediated coupling between quantum emitters, leading to collective atomic states such as super- and sub-radiant ensembles \cite{sheremet2023waveguide,tiranov2023collective,you2018waveguide,goban2015superradiance}. The spatial arrangement of emitters couple to a waveguide governs their radiative behavior, providing new avenues for controlling photon transport. Theoretical and experimental studies have extensively examined photon scattering in a one-dimensional waveguide \cite{shen2009theory,liao2015single,tsoi2008quantum,derouault2014one,cheng2017waveguide,konyk2017one,le2014propagation,huck2011controlled,yalla2014cavity,javadi2015single,li2016probing}, providing insights into transparency \cite{mukhopadhyay2020transparency,liao2016dynamical}, nonreciprocal photon dynamics \cite{miao2024kerr,wang2019nonreciprocity,fang2017multiple,rosario2018nonreciprocity,sounas2017non,cotrufo2021nonlinearity}, and nonclassical correlations \cite{claudon2010highly,zhang2018quantum}. For instance, periodic arrays of emitters coupled to waveguides can act as Bragg mirrors, enabling customized light reflection and transmission properties \cite{corzo2016large,sorensen2016coherent,corzo2019waveguide}. Chirality, combined with waveguide platforms, introduces directional photon-atom coupling, which significantly modifies relaxation rates into left- and right-propagating modes \cite{mitsch2014quantum,kornovan2017transport,mirza2017chirality}. This directional coupling facilitates dissipation-driven formation of entangled states among emitters, unlocking new possibilities for quantum technologies \cite{mirza2016multiqubit,mirza2016two,stannigel2012driven,gonzalez2015chiral,pichler2015quantum,ramos2014quantum}. Furthermore, advancements such as giant atoms in waveguide QED has revealed intriguing decoherence-free interactions and expanded the scope of quantum dynamics achievable within these systems \cite{kockum2018decoherence,kannan2020waveguide,wang2021tunable,wang2024realizing}.

Transparency in quantum systems has been an area of extensive investigation in the context of coherent manipulation. However, Mukhopadhyay and Agarwal discovered a different kind of transparency in the context of asymmetrically coupled qubits to a waveguide \cite{mukhopadhyay2020transparency}. Previous studies examined transparency in nonchiral waveguides under weak field conditions, attributing it to symmetry-driven interference effects between multiple quantum emitters. However, these results were often restricted to negligible input power, where transparency mimics classical analogs. In contrast, we demonstrate a robust form of transparency in a two-atom system coupled to a chiral waveguide with antisymmetric detunings. By introducing specific phase shifts, we achieve perfect transmission and zero reflectivity, independent of input power. This marks a significant departure from prior work confined to weak-field regimes, showcasing a transparency mechanism that persists even under strong-field conditions.

Nonclassical phenomena, such as photon antibunching and quantum squeezing, are foundational to advancements in quantum technologies. The hallmark of these phenomena lies in their deviation from classical photon statistics, exemplified by the second-order correlation function $g^{(2)}(0)$. While classical systems generate coherent fields with $g^{(2)}(0)=1$, two-level quantum systems introduce a mix of coherent and incoherent components, breaking the classical paradigm. Chiral waveguides, which asymmetrically couple quantum emitters to directional photonic modes, provide a versatile platform for generating nonclassical light. Our work builds on previous studies of photon antibunching and quantum correlations, investigating how phase shifts and detuning asymmetries govern the generation and control of nonclassical states. By tuning system parameters, such as coupling strengths and input fields, we reveal mechanisms for engineering strong photon antibunching and thus achieving high-efficiency single-photon sources.

Nonreciprocity, where light propagation is direction-dependent, is important in both classical and quantum photonics. In classical systems, nonreciprocity has been achieved using methods such as phase matching \cite{dong2015brillouin,kim2015non}, Kerr nonlinearities \cite{miao2024kerr}, and synthetic magnetic fields \cite{peterson2019strong,biehs2023enhancement}. However, translating these effects into the quantum regime remains a challenging frontier though Hamann et al. demonstrated strong nonreciprocity by especial choice of phase separation and detunings \cite{rosario2018nonreciprocity}. Chiral waveguides provide a promising platform for achieving quantum nonreciprocity by leveraging quantum nonlinear effects natural to qubits. In our study, we show that asymmetries in detunings and coupling strengths create conditions where transmission is suppressed in one direction while remaining finite in the other. Unlike classical systems, where nonreciprocity often relies on specific damping rates, the quantum regime offers directionality even in the absence of intrinsic damping. This tunable quantum nonreciprocity holds significant potential for applications in nonreciprocal quantum devices and scalable quantum networks.

In this paper, we present a unified theoretical framework that integrates these phenomena—transparency, nonclassicality, and nonreciprocity—into the broader context of chiral waveguide QED. Section II establishes connection between the temporal coupled-mode thory (TCMT) \cite{suh2004temporal,zhao2019connection} used extensively in classical optics with the master equation used in quantum optics. This connection is established via adding quantum Langevin forces to basic equations of TCMT. Section III explores the novel transparency effects in chiral waveguides, emphasizing their independence from input power. In Section IV, we analyze coherent and incoherent scattering phenomena, the latter being especially significant at higher driving powers. Section V focuses on engineering quantum light by manipulating qubit detunings and phase separations, with an emphasis on photon antibunching. Finally, Section VI examines quantum nonreciprocity, demonstrating how quantum nonlinear dynamics break reciprocity in light transmission and quantum fluctuations in amplitudes. Together, these results advance the understanding of chiral waveguide QED and highlight its potential for quantum photonic applications.

\section{From Temporal Coupled-Mode Theory to the Master Equation in Chiral Waveguide QED}
\begin{figure}[b]
	\includegraphics[width=8.6cm]{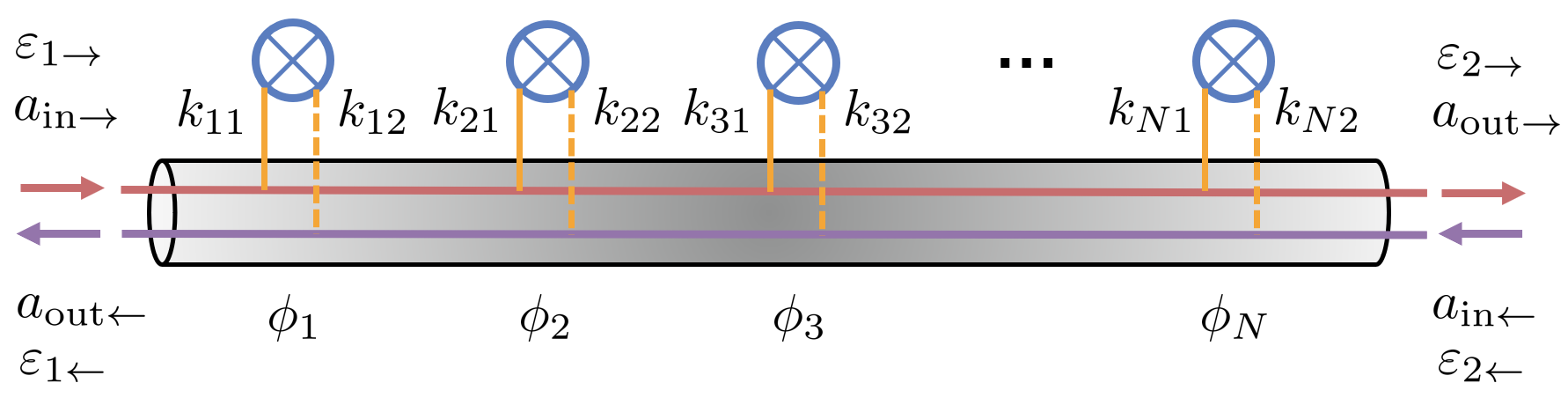}
	\caption{Schematic of a system with $N$ resonators coupled to incoming coherent fields $(\varepsilon_{1\rightarrow}, \varepsilon_{2\leftarrow})$ from two ports, with coupling strengths $(k_{i1},k_{i2})$, respectively. The output fields are $(\varepsilon_{1\leftarrow}, \varepsilon_{2\rightarrow})$. When treated quantum mechanically, the quantum vacuum input fields $(a_{\rm in\rightarrow}, a_{\rm in\leftarrow})$ and output fields $(a_{\rm out\leftarrow}, a_{\rm out\rightarrow})$ are also considered. The system’s chiral interaction with the reservoir is represented by the asymmetry in coupling strengths, $k_{i1} \neq k_{i2}$. $\phi_i$ is the phase shifts experienced by the forward propagating fields as they reach resonator $i$.}
	\label{model}
\end{figure}
\subsection{Temporal Coupled-Mode Theory}
We begin by considering an array of $N$ harmonic oscillators, each with a resonance frequency $\omega_i$ ($i = 1, 2, \dots, N$), coupled to the waveguide, as illustrated in Fig. \ref{model}. Each resonator can be coupled to the waveguide with different coupling strengths to the forward and backward propagating modes of the waveguide. The system Hamiltonian (we set $\hbar=1$),
\begin{equation}
	\begin{aligned}
		\hat H&=\sum_{i=1}^N\Delta_i c_i^\dagger c_i+i\sum_{i=1}^N(k_{i1}e^{i\phi_i}\varepsilon_{1\rightarrow} c_i^\dagger-k_{i1}^*e^{-i\phi_i} \varepsilon_{1\rightarrow}^*c_i)\\
		&+i\sum_{i=1}^N(k_{i2}e^{i(\phi_N-\phi_i)}\varepsilon_{2\leftarrow}  c_i^\dagger -k_{i2}^*e^{-i(\phi_N-\phi_i)} \varepsilon_{2\leftarrow}^* c_i),
	\end{aligned}
\end{equation}
is formulated in the rotating frame of the drive field, consisting of both the shifted energy terms of the resonators and their interactions with the drive field. The drive field comes from port $1$ or port $2$ -- one sending a forward propagating wave $(\varepsilon_{1\rightarrow})$ or a backward propagating wave $(\varepsilon_{2\leftarrow})$. Both of these fields are coherent, and operate at a frequency $\omega_d$. Each resonator $i$ has a detuning, $\Delta_i = \omega_i - \omega_d$. The resonators interact with the photons traveling through the waveguide, and the coupling is characterized by the coupling constants $k_{i\ell}$ where $\ell = 1,\ 2$ corresponds to the two ports. As the waves propagate, they undergo phase shifts. The phase shifts experienced by the forward and backward propagating fields as they reach resonator $i$ are $\phi_i$ and $\phi_N - \phi_i$, respectively. We can set $\phi_1=0$ without loss of generality.

Using a semiclassical approach and TCMT \cite{miao2024kerr}, the dynamics of the system in the rotating frame of the drive are described by
\begin{equation}
	\begin{aligned}
		\frac{d}{dt}\mathbf c(t)&=-i\mathbf\Delta\mathbf c(t)-\mathbf \Gamma\mathbf c(t)+\mathbf K\pmb \varepsilon_{\rm in},
	\end{aligned}
\end{equation}
where $\mathbf{c}(t) = (c_1(t), c_2(t), \dots, c_N(t))^T$, $\mathbf{\Delta} = \mathrm{diag}\{\Delta_1,\Delta_2, \dots, \Delta_N\}$, $\mathbf{\Gamma}$ is an $N \times N$ matrix capturing the decay processes, $\pmb{\varepsilon}_{\rm in} = (\varepsilon_{1\rightarrow}, \varepsilon_{2\leftarrow})^T$, and $\mathbf{K}$ is an $N \times 2$ matrix,
\begin{equation}
	\mathbf K=\left(\begin{array}{cc}
		k_{11} & k_{12}e^{i\phi_N} \\
		k_{21}e^{i\phi_2} & k_{22}e^{i(\phi_N-\phi_2)} \\
		\vdots & \vdots \\
		k_{i1}e^{i\phi_i} & k_{i2}e^{i(\phi_N-\phi_i)} \\
		\vdots & \vdots \\
		k_{N1}e^{i\phi_N} & k_{N2}
	\end{array}
	\right),
\end{equation}
determined by the structure of the system Hamiltonian. 

Applying the energy conservation constraint allows us to derive the relationship between $\mathbf K$ and $\mathbf\Gamma$ \cite{miao2024kerr}, 
\begin{equation}
	\mathbf K\mathbf K^\dagger=\mathbf \Gamma+\mathbf \Gamma^\dagger,
	\label{KG}
\end{equation}
along with the input-output relation,
\begin{equation}
	\pmb{\varepsilon}_{\rm out}=\mathbf C(\pmb{\varepsilon}_{\rm in}-\mathbf K^\dagger\mathbf c),
\end{equation} 
where $\pmb{\varepsilon}_{\rm out}=(\varepsilon_{1\leftarrow},\varepsilon_{2\rightarrow})^T$, $\mathbf C=\left(\begin{array}{cc}
	0 & e^{i\phi_N} \\
	e^{i\phi_N} & 0
\end{array}
\right)$ for the direct scattering in the absence of resonators. 

The waveguide acts as a mediator for the dissipative coupling between the resonators. Physically, $\Gamma_{ij}$ represents the coupling along the traveling wave from resonator $j$ to resonator $i$. For $i < j$, this coupling occurs via the backward-propagating field and is determined by the coupling strengths $k_{i2}$ and $k_{j2}$. Conversely, for $i > j$, the coupling occurs via the forward-propagating field and is determined by the coupling strengths $k_{i1}$ and $k_{j1}$. Thus, the elements of the matrix $\mathbf{\Gamma}$ can be explicitly determined. For the diagonal elements, we have
\begin{equation}
	\Gamma_{ii}=\frac{|k_{i1}|^2+|k_{i2}|^2}{2},
\end{equation}
and for the off-diagonal elements with $i < j$, 
\begin{equation}
	\Gamma_{ij}=k_{i2}k_{j2}^*e^{i(\phi_j-\phi_i)},\qquad\Gamma_{ji}=k_{i1}^*k_{j1}e^{i(\phi_j-\phi_i)}.
\end{equation}

\subsection{Transition from Semiclassical Coupled Mode Equations to Quantum Langevin Equations and the Master Equation}
When the system is treated quantum mechanically, $c_i$ should be regarded as bosonic operators. To preserve the commutation relation $[c_\alpha(t), c^\dagger_\beta(t)] = \delta_{\alpha\beta}$ for these operators, it is necessary to include operator Langevin forces with zero mean value leading to the quantum Langevin equations,
\begin{equation}
	\begin{aligned}
		\frac{d}{dt}\mathbf c(t)&=-i\mathbf\Delta\mathbf c(t)-\mathbf \Gamma\mathbf c(t)+\mathbf K(\pmb \varepsilon_{\rm in}+\mathbf a_{\rm in}(t)),
	\end{aligned}
	\label{qle}
\end{equation}
where the Langevin forces term, $\mathbf{a}_{\rm in}(t) = (a_{\rm in\rightarrow}(t), a_{\rm in\leftarrow}(t))^T$, also represents the quantum vacuum input fields.

For simplicity, we assume no input coherent field and no detunings, allowing Eq. (\ref{qle}) to be integrated as
\begin{equation} 
	\mathbf{c}(t) = e^{-\mathbf{\Gamma} t} \mathbf{c}(0) + \int_0^t e^{-\mathbf{\Gamma} (t - \tau)} \mathbf{K} \mathbf{a}_{\rm in}(\tau)d\tau.
\end{equation}
The commutator at time $t$ is given by
\begin{equation} 
	\begin{aligned}\
	[c_\alpha(t), c_\beta^\dagger(t)] &= [ e^{-\mathbf{\Gamma} t} e^{-\mathbf{\Gamma}^\dagger t}\\
	& + \int_0^t e^{-\mathbf{\Gamma} (t - \tau)} \mathbf{K} \mathbf{K}^\dagger e^{-\mathbf{\Gamma}^\dagger (t - \tau)}d\tau ]_{\alpha\beta},
	\end{aligned}
\end{equation}
where we use the commutation properties $[c_\alpha(0), c_\beta^\dagger(0)] = \delta_{\alpha\beta}$, $[a_{\rm in\alpha}(t_1), a_{\rm in\beta}^\dagger(t_2)] = \delta_{\alpha\beta} \delta(t_1 - t_2)$ and $[c_\alpha(0), a_{\rm in\beta}^\dagger(t)] =0$.

Following $[c_\alpha(t),c^\dagger_\beta(t)]=\delta_{\alpha\beta}$ and $d[c_\alpha(t),c^\dagger_\beta(t)]/dt=0$, and let $t'=t-\tau$, we arrive at the relationship $\mathbf K\mathbf K^\dagger=\mathbf \Gamma+\mathbf \Gamma^\dagger$ consistent with Eq. (\ref{KG}). This result validates our quantum Langevin equations.

We can construct the quantum master equation for the density matrix $\rho$ of the system from quantum Langevin equations Eq. (\ref{qle}), with the result
\begin{equation}
	\begin{aligned}
		\dot\rho&=-i[\hat H,\rho]-\sum_{i=1}^N(\Gamma_{ii}+\gamma_i)( c_i^\dagger c_i\rho-2 c_i\rho c_i^\dagger+\rho c_i^\dagger c_i)\\
		&\!\!\!\!\!\!\!\!\!-\sum_{i<j}^N[\Gamma_{ij} c_i^\dagger c_j\rho-(\Gamma_{ij}+\Gamma_{ji}^*) c_j\rho c_i^\dagger+\Gamma_{ji}^* \rho c_i^\dagger c_j+H.c.],
	\end{aligned}
\label{bm}
\end{equation}
where we include the intrinsic damping rate $\gamma_i$ of each resonator. 

\subsection{Quantum Master Equation for Two-Level Atoms}
In this section, we present the general form of the quantum master equation for an array of $N$ two-level atoms coupled to a one-dimensional chiral waveguide. For two-level atoms, the master equation Eq. (\ref{bm}) can be adapted by replacing the resonator annihilation and creation operators, $c_i$ and $c_i^\dagger$, with the atomic lowering and raising operators, $S_i^-$ and $S_i^+$ for each atom $i$, respectively. This is because the structure of master equation is same [see Eq. (9.28) in \cite{agarwal2012quantum}]. Consequently, the most general master equation for an array of $N$ two-level atoms coupled to a chiral waveguide is given by
\begin{equation}
	\begin{aligned}
		\dot\rho&=-i[\hat H,\rho]-\sum_{i=1}^N(\Gamma_{ii}+\gamma_i)( S_i^+ S_i^-\rho-2 S_i^-\rho S_i^++\rho S_i^+ S_i^-)\\
		&\!\!\!\!\!\!\!\!\!-\sum_{i<j}^N[\Gamma_{ij} S_i^+ S_j^-\rho-(\Gamma_{ij}+\Gamma_{ji}^*) S_j^-\rho S_i^++\Gamma_{ji}^* \rho S_i^+ S_j^-+H.c.],
	\end{aligned}
\end{equation}
with the system Hamiltonian
\begin{equation}
	\begin{aligned}
		\hat H&=\sum_{i=1}^N\Delta_i S_i^+ S_i^-+i\sum_{i=1}^N(k_{i1}e^{i\phi_i}\varepsilon_{1\rightarrow} S_i^+-k_{i1}^*e^{-i\phi_i}\varepsilon_{1\rightarrow}^* S_i^-)\\
		&+i\sum_{i=1}^N(k_{i2}e^{i(\phi_N-\phi_i)}\varepsilon_{2\leftarrow}  S_i^+ -k_{i2}^*e^{-i(\phi_N-\phi_i)}\varepsilon_{2\leftarrow}^*  S_i^-),
	\end{aligned}
\end{equation}
and input-output relations
\begin{equation}
	\begin{aligned}
		&\varepsilon_{1\leftarrow}=e^{i\phi_N}\varepsilon_{2\leftarrow}-\sum_{i=1}^Nk_{i2}^*e^{i\phi_i}S_i^-,\\
		&\varepsilon_{2\rightarrow}=e^{i\phi_N}\varepsilon_{1\rightarrow}-\sum_{i=1}^Nk_{i1}^*e^{i(\phi_N-\phi_i)}S_i^-.
	\end{aligned}
\end{equation} 
In writing input-output relations, we drop vacuum terms as these do not contribute to normally ordered expectation values. Note that for real $k_{i1}$ and $k_{i2}$, the signs may differ, as observed in magnetic transitions \cite{wang2019nonreciprocity}. If all coupling strengths are positive, the master equation reduces to the form presented by Pichler et al. \cite{pichler2015quantum}. Our derivation links the quantum master equation and the quantum Langevin equations to the classical TCMT through the inclusion of quantum vacuum field terms.

\section{Transparency in a Two-Atom Waveguide QED with Antisymmetric Detunings}
\label{Trans}

In a previous work by Mukhopadhyay et al. \cite{mukhopadhyay2020transparency}, transparent behavior was observed in a chain of quantum emitters coupled to a nonchiral waveguide, limited to weak input fields. In contrast, here we extend this phenomenon, demonstrating transparency in a two-atom system for all input field strengths and under chiral coupling conditions with the waveguide. Consider a system where two atoms, $a$ and $b$, are coupled to a waveguide with the same coupling constants for each port: $k_{a1} = k_{b1}=k_1$ and $k_{a2} = k_{b2}=k_2$. However, the detunings are antisymmetric -- while atom $a$ is offset by $\Delta$, atom $b$ is offset by $-\Delta$ ($\Delta_a=-\Delta_b=\Delta$). This antisymmetry, paired with the phase shift $\phi=n\pi$ ($n = 1, 2, \dots$), allows us to achieve perfect transparency, characterized by complete transmission and zero reflectivity, across all input field strengths in an ideal system with no intrinsic damping at all ($\gamma_a = \gamma_b = \gamma = 0$). 

We turn off the backward drive, setting $\varepsilon_{2\leftarrow} = 0$, restricting the drive to forward propagation. In this case, the system Hamiltonian is formulated as
\begin{equation}
	\begin{aligned}
	\hat H&=\Delta_a S_a^z+\Delta_b S_b^z\\
	&+ik_{1}\alpha S_a^+-ik_{1}^*\alpha^*S_a^-+ik_{1}e^{i\phi}\alpha S_b^+-ik_{1}^*e^{-i\phi}\alpha^*S_b^-,
	\end{aligned}
\end{equation}
where we denote the complex amplitude of the forward-propagating field as $\alpha$ ($\varepsilon_{1\rightarrow} = \alpha$). The input-output relations are
\begin{equation}
	\begin{aligned}
		&\varepsilon_{1\leftarrow}=-k_{2}^*(S_a^-+e^{i\phi}S_b^-),\\
		&\varepsilon_{2\rightarrow}=e^{i\phi}\alpha-k_{1}^*(e^{i\phi}S_a^-+S_b^-).
	\end{aligned}
\label{2io}
\end{equation} 

We now examine the expectation values in the long-time limit ($\dot{\rho} = 0$), using the following state basis: the fully excited state $|e\rangle = |e_a, e_b\rangle$, the ground state $|g\rangle = |g_a, g_b\rangle$, the symmetric triplet state $|s\rangle = (|e_a, g_b\rangle + |g_a, e_b\rangle)/\sqrt{2}$, and the antisymmetric singlet state $|a\rangle = (|e_a, g_b\rangle - |g_a, e_b\rangle)/\sqrt{2}$. Here, we consider the phase shift $\phi$ to be an integer multiple of $\pi$ (i.e., $\phi = n\pi$ for $n = 1, 2, \dots$). For odd values of $n$, i.e., $n=2k-1$ ($k=1,2,\dots$), the expectation value of the output field operator $\varepsilon_{2\rightarrow}$ is
\begin{equation}
	\langle\varepsilon_{2\rightarrow}\rangle=-\alpha-\sqrt2k_1^*(\rho_{ea}-\rho_{ag}),
\end{equation} 
and the expectation value of the photon number operator $\varepsilon_{2\rightarrow}^\dagger \varepsilon_{2\rightarrow}$ is
\begin{equation}
	\begin{aligned}
		\langle\varepsilon_{2\rightarrow}^\dagger \varepsilon_{2\rightarrow}\rangle
		&=|\alpha|^2+|k_{1}|^2(2\rho_{ee}+2\rho_{aa})\\
		&-\sqrt2[k_{1}^*\alpha^*(-\rho_{ea}+\rho_{ag})+H.c.].
	\end{aligned}
\end{equation} 
In an ideal system with antisymmetric detunings and no intrinsic damping, the two-atom system settles into a pure quantum state in the long-time limit when $\phi = n\pi$ \cite{pichler2015quantum}. For odd values of $n$, the density matrix reveals that the system occupies a superposition of the triplet state and the ground state, expressed as $\rho = |\psi\rangle\langle\psi|$, where $|\psi\rangle = a|s\rangle + b|g\rangle$. Thus
\begin{equation}
	\rho_{ee}=\rho_{ea}=\rho_{ag}=0.
\end{equation} 
This implies that not only is the amplitude of the transmitted field equal to that of the incident field, but the mean intensity of the output equals the input intensity as well.

\begin{figure}[b]
	\includegraphics[width=9cm]{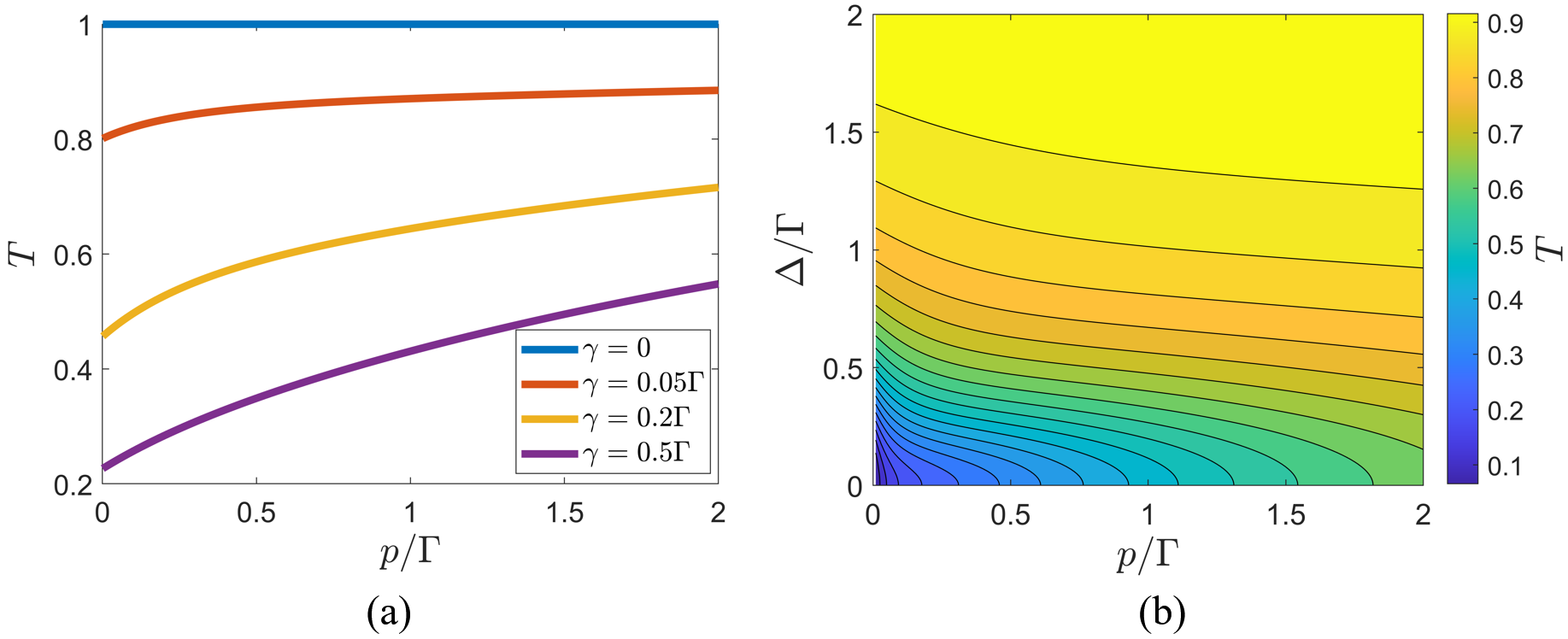}
	\caption{(a) Power transmission $T$ as a function of input power $p$ (we define $p=\alpha^2$ and set $\alpha$ real) for various values of intrinsic damping, $\gamma_a = \gamma_b = \gamma$. The parameters used are $\Delta_a = -\Delta_b = \Delta = \Gamma$, $k_{a1} = k_{b1} = k_1 = \sqrt{1.2\Gamma}$, $k_{a2} = k_{b2} = k_2 = \sqrt{0.8\Gamma}$, and $\phi = \pi$. (b) Contour plot of power transmission $T$ as a function of input power $p$ and detuning $\Delta$. The parameters are $\gamma = 0.05\Gamma$, $k_{a1} = k_{b1} = k_1 = \sqrt{1.2\Gamma}$, $k_{a2} = k_{b2} = k_2 = \sqrt{0.8\Gamma}$, and $\phi = \pi$.}
	\label{transparency}
\end{figure}

For even values of $n$, i.e., $n=2k$ ($k=1,2,\dots$), the expectation values become
\begin{equation}
	\langle\varepsilon_{2\rightarrow}\rangle=\alpha-\sqrt2k_1^*(\rho_{es}+\rho_{sg}),
\end{equation} 
and
\begin{equation}
	\begin{aligned}
		\langle\varepsilon_{2\rightarrow}^\dagger \varepsilon_{2\rightarrow}\rangle
		&=|\alpha|^2+|k_{1}|^2(2\rho_{ee}+2\rho_{ss})\\
		&-\sqrt2[k_{1}^*\alpha^*(\rho_{es}+\rho_{sg})+H.c.].
	\end{aligned}
\end{equation} 
In this case, the system occupies in a superposition of the singlet state and the ground state, expressed as $|\psi\rangle = a|a\rangle + b|g\rangle$. Thus
\begin{equation}
	\rho_{ee}=\rho_{es}=\rho_{sg}=0.
\end{equation}
As a result, the output field reduces to $\langle \varepsilon_{2\rightarrow} \rangle = \pm\alpha$ (with $+$ for even $n$ and $-$ for odd $n$), and the output photon numbers remains $\langle \varepsilon_{2\rightarrow}^\dagger \varepsilon_{2\rightarrow} \rangle = |\alpha|^2$ in both cases.

In the even $n$ case, the nonzero density matrix elements $\rho_{aa}$ and $\rho_{ag}$ mirror $\rho_{ss}$ and $\rho_{sg}$ from the odd $n$ case, respectively,
\begin{equation}
	\begin{aligned}
	\rho_{ss}|_{n=2k-1}&=\rho_{aa}|_{n=2k}\\
	&\!\!\!\!\!\!=\frac{8|k_1|^2|\alpha|^2}{(|k_1|^2-|k_2|^2)^2+8|k_1|^2|\alpha|^2+4\Delta^2},
	\end{aligned}
\end{equation}
\begin{equation}
	\begin{aligned}
	\rho_{sg}|_{n=2k-1}&=\rho_{ag}|_{n=2k}\\
	&\!\!\!\!\!\!=\frac{2\sqrt2k_1\alpha(|k_1|^2-|k_2|^2-2i\Delta)}{(|k_1|^2-|k_2|^2)^2+8|k_1|^2|\alpha|^2+4\Delta^2}.
	\end{aligned}
\end{equation}
This result is of importance in stabilizing remote entanglement between a pair of noninteracting superconducting qubits via waveguide-mediated dissipation, as demonstrated in a recent experiment \cite{shah2024stabilizing}.

In our system, we conclude that it maintains complete transmission, regardless of the input power. The field amplitude transmission is $|t| = |\langle \varepsilon_{2\rightarrow} \rangle / \alpha| = 1$, and the power transmission is $T = \langle \varepsilon_{2\rightarrow}^\dagger \varepsilon_{2\rightarrow} \rangle / |\alpha|^2 = 1$, confirming perfect transparency. 

When intrinsic damping is introduced into the two-atom system, the steady system begins to deviate from its ideal pure quantum state, leading to a reduction in the power transmission $T$. As shown in Fig. \ref{transparency}(a), the transmission decreases as intrinsic damping increases. However, when the damping is set to $\gamma_a = \gamma_b = \gamma=0.05\Gamma$, where $\Gamma = (|k_1|^2 + |k_2|^2)/2$, the transmission remains notably high for antisymmetric detunings $\Delta_a = -\Delta_b = \Delta = \Gamma$. Fig. \ref{transparency}(b) further illustrates the power transmission $T$ becomes more pronounced as both the detuning $\Delta$ and input power $p$ increase when $\gamma=0.05\Gamma$. 

\section{Coherent and Incoherent Scattering in a chiral waveguide}
\label{fluc}

\begin{figure}[b]
	\includegraphics[width=9cm]{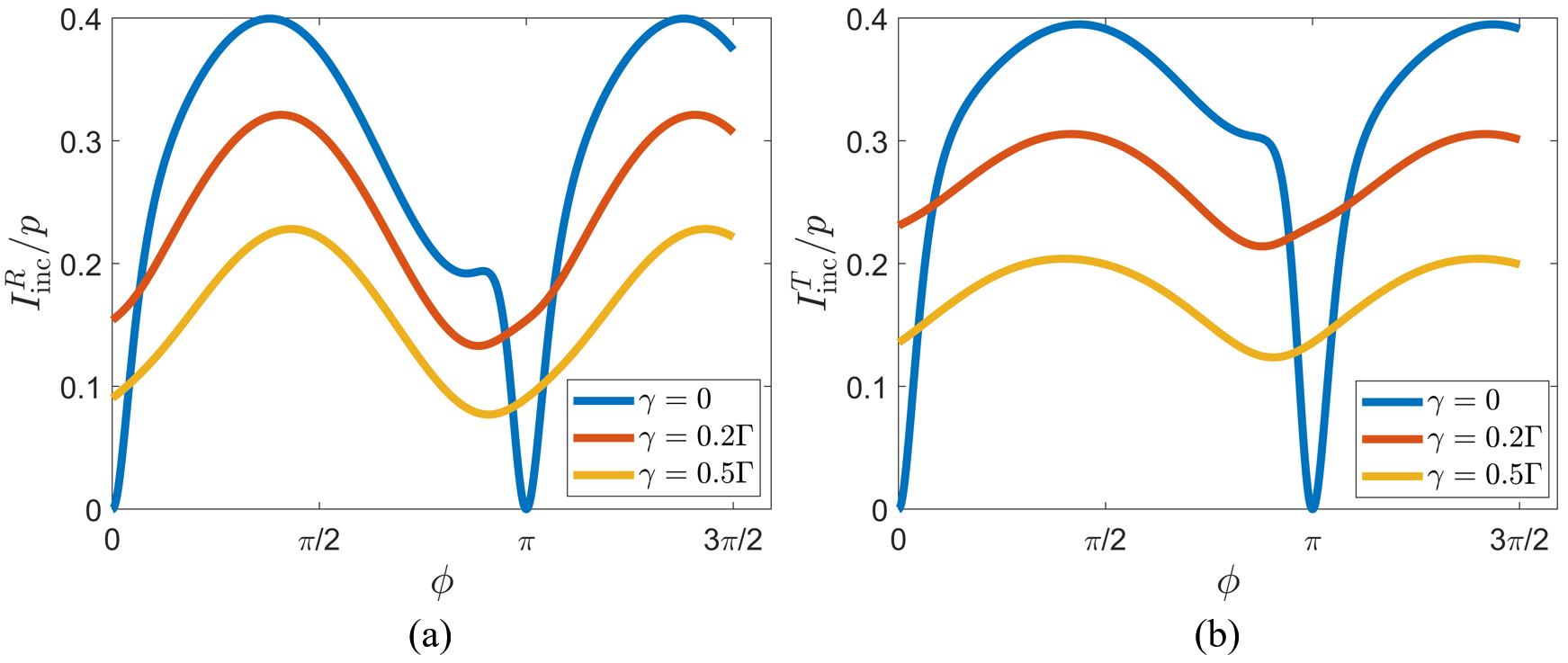}
	\caption{The incoherent component of the output for (a) the reflected field, $I_{\rm inc}^R/p$, and (b) the transmitted field, $I_{\rm inc}^T/p$, as functions of the phase shift $\phi$, plotted for various intrinsic damping values $\gamma_a = \gamma_b = \gamma = 0$, $0.2\Gamma$, and $0.5\Gamma$. The parameters used are $\Delta_a = -\Delta_b = \Delta = 0.5\Gamma$, $k_{a1} = k_{b1} = k_1 = \sqrt{1.2\Gamma}$, $k_{a2} = k_{b2} = k_2 = \sqrt{0.8\Gamma}$, and input power $p = \Gamma$.}
	\label{I_inc_phi}
\end{figure}
\begin{figure}[b]
	\includegraphics[width=9cm]{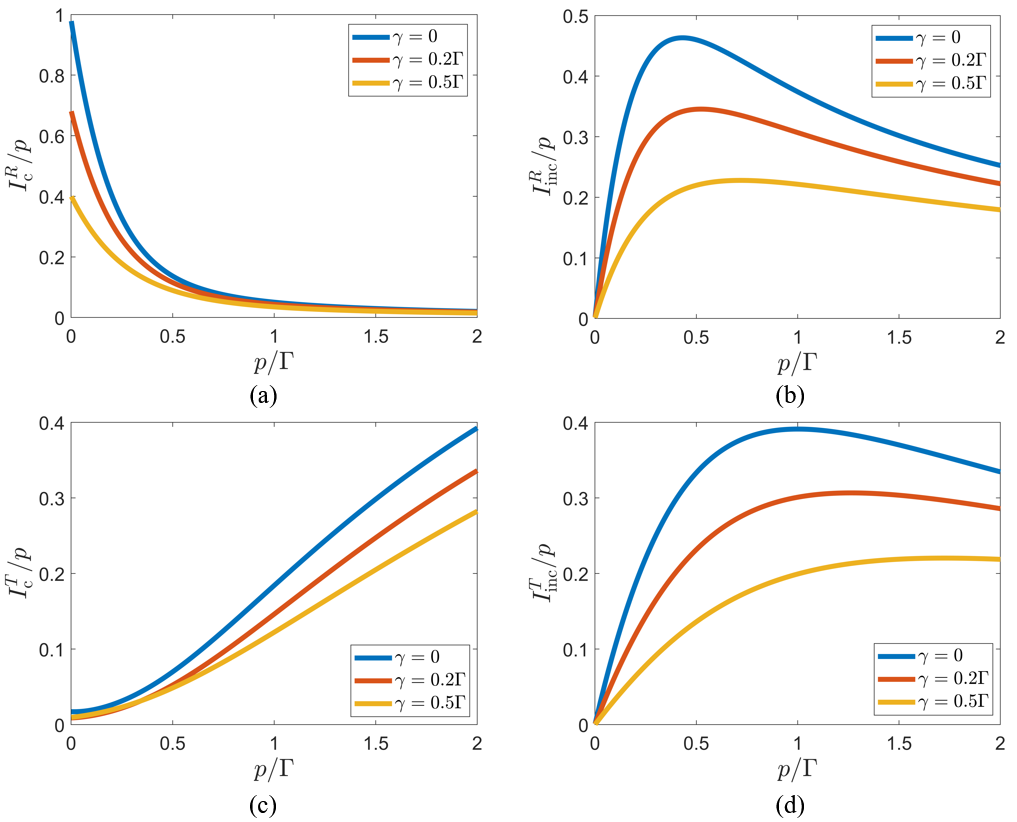}
	\caption{The coherent component ($I_{\rm c}/p$) and incoherent component ($I_{\rm inc}/p$) of the output power for (a) and (b) the reflected field, and (c) and (d) the transmitted field, plotted as functions of the input power $p$ for various intrinsic damping values: $\gamma_a = \gamma_b = \gamma = 0$, $0.2\Gamma$, and $0.5\Gamma$. The parameters used are $\Delta_a = -\Delta_b = \Delta = 0.5\Gamma$, $k_{a1} = k_{b1} = k_1 = \sqrt{1.2\Gamma}$, $k_{a2} = k_{b2} = k_2 = \sqrt{0.8\Gamma}$, and $\phi = \pi/2$.}
	\label{fluctuations}
\end{figure}
In this section and the next, we explore the quantum features of a two-atom system with antisymmetric detunings coupled to a chiral waveguide. Here, we focus on coherent and incoherent scattering phenomena for phase separation different from an integeral multiple of $\pi$. The intensity of the coherent component of the output fields can be quantified as 
\begin{equation}
	I_{\rm c}^R=|\langle \varepsilon_{1\leftarrow}\rangle|^2,\qquad I_{\rm c}^T=|\langle \varepsilon_{2\rightarrow}\rangle|^2,
\end{equation}
corresponding to the coherent contributions to the reflected and transmitted fields, respectively. The total output powers for reflection and transmission are given by
\begin{equation}
	I^R=\langle \varepsilon_{1\leftarrow}^\dagger\varepsilon_{1\leftarrow}\rangle,\qquad I^T=\langle \varepsilon_{2\rightarrow}^\dagger\varepsilon_{2\rightarrow}\rangle,
\end{equation}
and the net incoherent part of the output power is determined by subtracting the coherent component from the total intensity,
\begin{equation}
	I_{\rm inc}^R=I^R - I_{\rm c}^R,\qquad I_{\rm inc}^T=I^T - I_{\rm c}^T,
\end{equation}
which can arise from quantum nonlinearities inherent in the system. In a chiral waveguide system with classical harmonic oscillators, coherent excitation leads to purely coherent radiation. This behavior arises because harmonic oscillators possess an infinite number of equally spaced energy levels, allowing them to respond linearly to a coherent drive. Quantum fluctuations in the output field amplitudes vanish, such that $I_{\rm inc}^R=0$, $I_{\rm inc}^T=0$, regardless of the input power. 

For coherently driven two-level atoms, the situation is markedly different due to the finite energy-level structure and the saturation effects. However, at very low input power ($p \rightarrow 0$), quantum nonlinear interactions are negligible, and the system behaves almost like its classical counterpart. In this regime, the coherent component dominates, with the incoherent part of the output power vanishes. As the input power increases, quantum nonlinear effects become prominent. 

As shown in Fig. \ref{I_inc_phi}, the incoherent part of the output power exhibits a periodic dependence on the phase shift $\phi$ with a period of $\pi$. The incoherent scattering is particularly prominent when $\phi$ is near $(2k - 1)\pi/2$ ($k = 1, 2, \dots$), far from the transparency condition, $\phi = n\pi$ ($n = 0, 1, 2, \dots$). This demonstrates the significant influence of the phase shift on quantum features in the chiral waveguide system. When the intrinsic damping $\gamma = 0$, the incoherent components vanish at $\phi = n\pi$, consistent with the transparency results discussed in Section \ref{Trans}. In Fig. \ref{fluctuations}, we fix the phase shift at $\phi = \pi/2$ and investigate how $I_{\rm c}/p$ and $I_{\rm inc}/p$ varies with the input power $p$ for both reflected and transmitted fields. The numerical results confirm that at low $p$, the coherent component dominates the output, with minimal contributions from incoherent scattering. As the input power increases, however, $I_{\rm inc}/p$ becomes significant, reaching a maximum at certain power levels. For high input powers, the behavior diverges between the reflected and transmitted fields. For the reflected field, $I_{\rm c}^R/p$ remains small and nearly unchanged. In contrast, for the transmitted field, the coherent component grows dominant as the input power increases. This behavior arises from the saturation of the two atoms at high power, which limits their nonlinear response and consequently diminishes the contribution of quantum fluctuations.

\section{Antibunching of fields produced by two atoms coupled to a chiral waveguide}
\label{antib}

In Section \ref{fluc}, we explored the quantum fluctuations (incoherent component) of output field amplitudes in a two-level atom system coupled to a chiral waveguide. In the low-input-power regime, these fluctuations become negligible, and the system closely resembles classical behavior. However, the second-order correlation function $g^{(2)}(0)$ deviates from this classical analogy. While quantum fluctuations in amplitude become negligible, $g^{(2)}(0)-1$, which quantifies fluctuations in intensity, can remain nonzero. For a single two-level atom coupled to a chiral waveguide, the reflected field satisfies $g^{(2)}_R(0) = 0$, regardless of the input power. This result arises from the quantum mechanical restriction that a two-level atom cannot emit two photons simultaneously due to its finite energy structure.

Extending our analysis to a two-atom system, we uncover enhanced opportunities for controlling photon antibunching in the reflected field, paving the way for the development of efficient single-photon sources. In the low-power regime, the reflected field exhibits sub-Poissonian second-order correlations, characterized by $g^{(2)}_R(0)$ dropping below 1 and approaching 0. The interaction between the two atoms and the backward-propagating wave modifies the emission dynamics, leading to a nonzero $g^{(2)}_R(0)$. Unlike the transmitted field, the reflected field lacks a direct contribution from the input coherent field (as described in Eq. (\ref{2io})), making it more appropriate for studies of photon antibunching and nonclassical light behavior. 

\begin{figure}[b]
	\includegraphics[width=9cm]{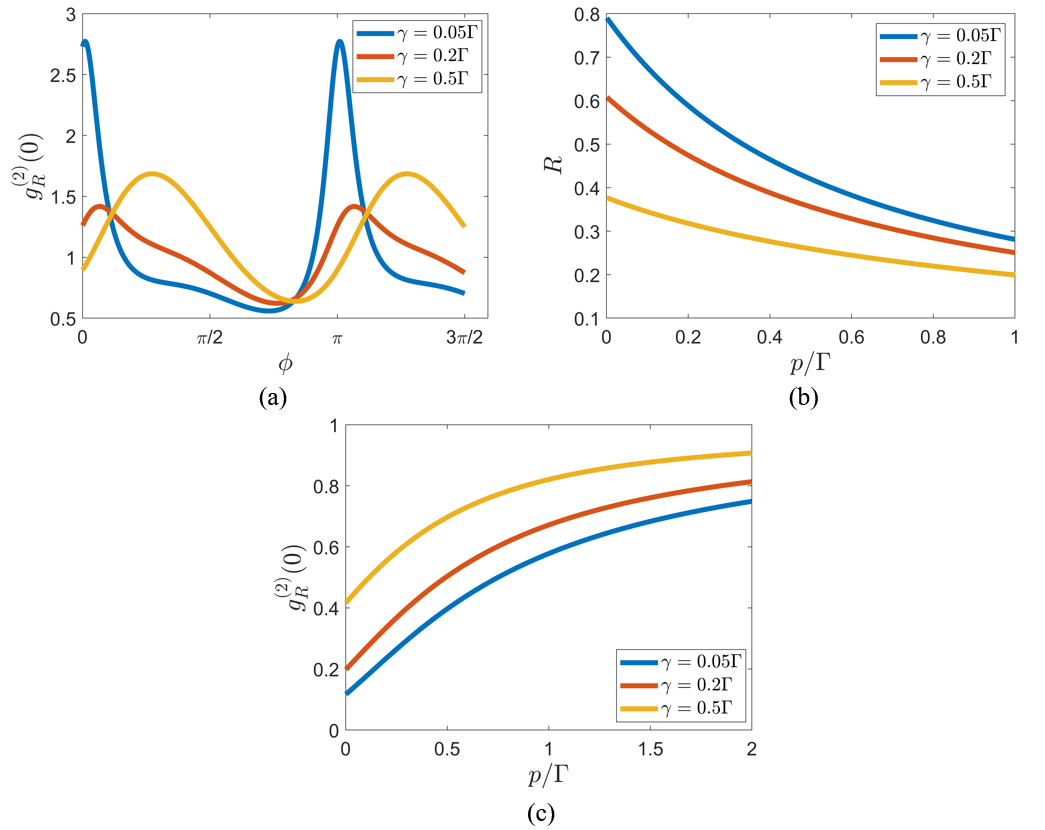}
	\caption{The second-order correlation function $g^{(2)}_R(0)$ of the reflection field and power reflectivity $R$ for various intrinsic damping values, $\gamma_a = \gamma_b = \gamma = 0.05\Gamma,\ 0.2\Gamma,\ 0.5\Gamma$. (a) The second-order correlation function $g^{(2)}_R(0)$ for the reflection field as a function of phase shift $\phi$, with parameters $\Delta_a = -\Delta_b = \Delta = 0.5\Gamma$, $k_{a1} = k_{b1} = k_1 = \sqrt{1.6\Gamma}$, $k_{a2} = k_{b2} = k_2 = \sqrt{0.4\Gamma}$, and input power $p = \Gamma$. (b) Power reflectivity $R$ and (c) $g^{(2)}_R(0)$ as functions of input power $p$, with parameters $\Delta_a = -\Delta_b = \Delta = 0.5\Gamma$, $k_{a1} = k_{b1} = k_1 = \sqrt{1.6\Gamma}$, $k_{a2} = k_{b2} = k_2 = \sqrt{0.4\Gamma}$, and $\phi = 2\pi/3$.}
	\label{g2}
\end{figure}

To quantify this behavior, we consider the system under specific parameters: the phase shift $\phi$ is fixed at $\pi/2$, and as in Section \ref{Trans}, $k_{a1} = k_{b1}=k_1$, $k_{a2} = k_{b2}=k_2$, and $\Delta_a=-\Delta_b=\Delta$ with $\gamma_a = \gamma_b = \gamma = 0$. In the low-power regime, the second-order correlation function for the reflected field can be approximated by expanding density matrix elements in $\alpha$,
\begin{equation}
	\begin{aligned}
	g^{(2)}_R(0)&=\frac{\langle\varepsilon_{1\leftarrow}^\dagger\varepsilon_{1\leftarrow}^\dagger \varepsilon_{1\leftarrow}\varepsilon_{1\leftarrow}\rangle}{|\langle\varepsilon_{1\leftarrow}^\dagger \varepsilon_{1\leftarrow}\rangle|^2}\\
	&\simeq\frac{|k_2|^4(|k_1|^4+6|k_1|^2|k_2|^2+|k_2|^4+4\Delta^2)^2}{[(|k_1|^2+|k_2|^2)^3+4(|k_1|^2+|k_2|^2)\Delta^2]^2}.
	\end{aligned}
\end{equation}
Additionally, the power reflectivity is approximated as
\begin{equation}
	\begin{aligned}
		R&=\frac{\langle\varepsilon_{1\leftarrow}^\dagger \varepsilon_{1\leftarrow}\rangle}{|\alpha|^2}\\
		&=\frac{16|k_1|^2|k_2|^2
			(|k_1|^4+2|k_1|^2|k_2|^2+|k_2|^4+4\Delta^2)}{(|k_1|^4+6|k_1|^2|k_2|^2+|k_2|^4+4\Delta^2)^2}.
		\end{aligned}
\end{equation}
For symmetric coupling ($k_1 = k_2$), the minimum value of $g^{(2)}_R(0)$ achievable is $1/4$, which occurs as $\Delta \to \infty$. However, the chirality of the waveguide provides additional flexibility through the asymmetry in $k_1$ and $k_2$, enabling stronger antibunching. For instance, with parameters $\Delta = 0.5\Gamma$, $k_1 = \sqrt{1.6\Gamma}$, and $k_2 = \sqrt{0.4\Gamma}$, we find the correlation function drops to $g^{(2)}_R(0) \simeq 0.09$. Simultaneously, the reflectivity is $R \simeq 0.9$, demonstrating the system's ability to achieve high-efficiency single-photon emission.

In Fig. \ref{g2}(a), the second-order correlation function $g^{(2)}_R(0)$ for the reflected field exhibits periodicity with a period of $\pi$, similar to the behavior observed for quantum fluctuations. The minimum value of $g^{(2)}_R(0)$ occurs near $\phi = 2\pi/3 + n\pi$ ($n = 0, 1, 2, \dots$), indicating strong photon antibunching at these phase shifts. Figures \ref{g2}(b) and \ref{g2}(c) show the power reflectivity $R$ and the corresponding $g^{(2)}_R(0)$ as functions of the input power $p$ at $\phi=2\pi/3$. The results demonstrate the system's capability to achieve sub-Poissonian statistics with $g^{(2)}_R(0) < 1$ while maintaining high reflectivity. This combination highlights the system's efficiency in converting classical coherent light into highly nonclassical quantum light.

\section{Quantum Nonreciprocity in chiral waveguied QED}

\begin{figure}[b]
	\includegraphics[width=9cm]{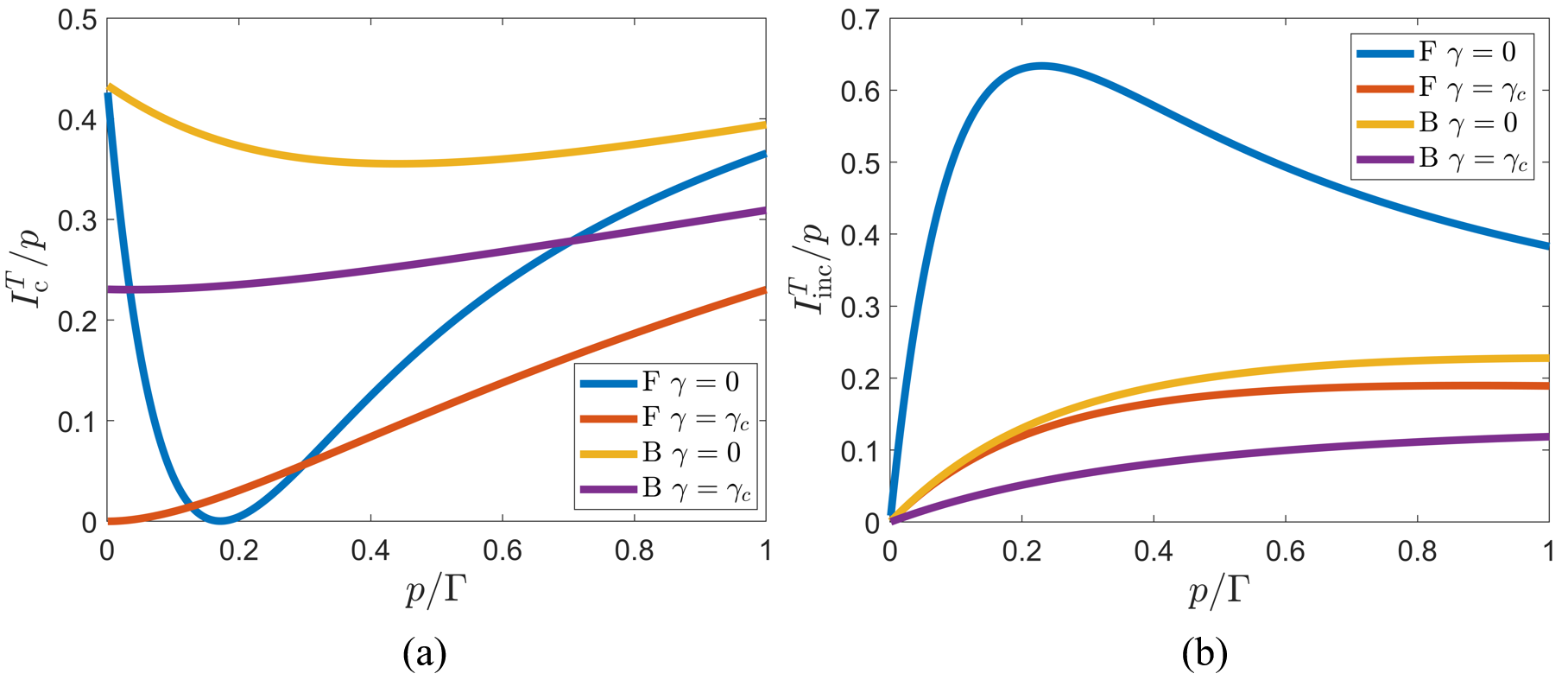}
	\caption{(a) The coherent component $I_{\rm c}^T/p$ and (b) the incoherent component $I_{\rm inc}^T/p$ of the transmitted field as functions of the input power $p$, plotted for various intrinsic damping values, $\gamma_a = \gamma_b = \gamma = 0,\gamma_c$ ($\gamma_c = 0.6\Gamma$), for both forward ($F$) and backward ($B$) propagation. The parameters are $\Delta_a = 0$, $\Delta_b = \Gamma$, $k_{a1} = k_{b2} = k_\alpha = \sqrt{1.6\Gamma}$, $k_{b1} = k_{a2} = k_\beta = \sqrt{0.4\Gamma}$, and $\phi = \pi$. }
	\label{nonreciprocity}
\end{figure}
Nonreciprocity in field amplitude transmission occurs when the forward transmission, defined as $|t_\rightarrow|=|\varepsilon_{2\rightarrow}/\alpha|$ (with $\varepsilon_{1\rightarrow}\neq0$, $\varepsilon_{2\leftarrow}=0$), differs from the backward transmission, $|t_\leftarrow|=|\varepsilon_{1\leftarrow}/\alpha|$ (with $\varepsilon_{1\rightarrow}=0$, $\varepsilon_{2\leftarrow}\neq0$). This phenomenon has been previously experimentally studied by Hamann et al. \cite{rosario2018nonreciprocity}, who demonstrated nonreciprocal behavior nonreciprocal behavior emerging from the interaction between two qubits embedded in a nonchiral waveguide, with quantum nonlinearity and asymmetric detuning used to break the structural symmetry of the system. However, with a nonchiral waveguide, the potential for nonreciprocity was inherently limited.

In chiral waveguide QED, we uncover a new type of quantum criticality that enables complete suppression of forward-propagating amplitude transmission at specific driving powers. The details of this new type of quantum criticality for a single two-level atom is discussed in Appendix \ref{appendix}. This phenomenon allows for an exceptionally high degree of nonreciprocity in both the coherent and incoherent components of the transmitted field. 

Extending this analysis to a two-atom system, we consider a parameter configuration where $k_{a1} = k_{b2} = k_\alpha$ and $k_{b1} = k_{a2} = k_\beta$, preserving structural symmetry between the two atoms. Under this setup, when the system consists of classical harmonic oscillators coupled to the waveguide without intrinsic damping, transmission is entirely reciprocal, regardless of the detunings $\Delta_a$ and $\Delta_b$ \cite{miao2024kerr}. The field amplitude transmission in this classical case is given by
\begin{equation}
	\begin{aligned}
	|t_\rightarrow|^2&=|t_\leftarrow|^2\\
	&\!\!\!\!\!\!=\frac{[\Delta_a^2+(|k_\alpha|^2-|k_\beta|^2)^2/4][\Delta_b^2+(|k_\alpha|^2-|k_\beta|^2)^2/4]}{|\Gamma_{a\rightarrow b}\Gamma_{b\rightarrow a}-(i\Delta_a+\Gamma)(i\Delta_b+\Gamma)|^2},
	\end{aligned}
\end{equation}
where $\Gamma=(|k_\alpha|^2+|k_\beta|^2)/2$, $\Gamma_{a\rightarrow b}=k_{a1}^*k_{b1}e^{i\phi}$, $\Gamma_{b\rightarrow a}=k_{a2}k_{b2}^*e^{i\phi}$. 

In contrast, in the quantum system with two two-level atoms, asymmetric detunings break reciprocity, revealing distinctive quantum nonlinear effects. Specifically, when $k_{\alpha} > k_{\beta}$, $\Delta_a = 0$, and $\Delta_b\neq0$, the system demonstrates a striking difference between forward and backward propagation. For forward propagation, the system can enter a quantum critical regime where the coherent field amplitude transmission vanishes entirely at specific driving powers, while the backward transmission remains finite. Unlike classical systems, where achieving critical coupling requires the intrinsic damping rate $\gamma_a$ to match a critical value $\gamma_c = (|k_{a1}|^2 - |k_{a2}|^2)/2$, complete suppression of transmission can occur even when $\gamma_a = 0$.

Figure \ref{nonreciprocity}(a) illustrates the coherent component of the transmitted field under different intrinsic damping conditions. For $\gamma = 0$, nonreciprocity is minimal at very low input power, as the system dynamics closely resemble the classical regime with negligible quantum nonlinear effects. As the input power increases, a transmission dip occurs near $p = 0.2\Gamma$ for forward propagation, marking a strong nonreciprocal response. When $\gamma = \gamma_c$, the transmission zero point shifts to $p = 0$, corresponding to classical critical coupling. In Fig. \ref{nonreciprocity}(b), we investigate the quantum fluctuation (incoherent component) of the transmitted field amplitude. For $\gamma = 0$, the forward-propagating field exhibits significantly larger quantum fluctuations compared to the backward-propagating field, amplifying the nonreciprocal response. However, when $\gamma = \gamma_c$, the nonreciprocity in quantum fluctuations diminishes.

\section{Conclusions} 
In this work, we have explored transparency, nonclassical light generation, and enhanced quantum nonreciprocity in chiral waveguide quantum electrodynamics (QED). For a two-atom system with antisymmetric detunings, we demonstrated complete forward transmission and zero reflectivity, irrespective of input power, achieved through phase separations of $\phi = n\pi$. Beyond the transparency condition, we investigated the coherent and incoherent components of the scattered fields for $\phi \neq n\pi$, and revealed strong photon antibunching in the reflected field as quantified by the second-order correlation function $g^{(2)}(0)$. By tuning system parameters such as detunings and coupling asymmetries, we demonstrated the capability to engineer single-photon sources with high efficiency and controllability, highlighting the versatility of chiral waveguide systems for quantum light generation. Furthermore, leveraging the intrinsic quantum nonlinearity of the system, we uncovered a novel type of quantum criticality that enables exceptional nonreciprocity in field amplitude transmission and amplitude fluctuations. Specifically, asymmetric detunings and coupling strengths were shown to induce direction-dependent transmission, including complete suppression in one propagation direction under certain conditions. These findings highlight the unique versatility and potential of chiral waveguide QED for quantum information and nonreciprocal quantum devices.

\section{Acknowledgments}
We are grateful for the support of Air Force Office of Scientific Research (Award No FA-9550-20-1-0366) and the Robert A Welch Foundation (Grants No. A-1943-20210327 and No. A-1943-20240404).

\appendix

\section{A New Type of Quantum Criticality for a Single Two-Level Atom}
\label{appendix}

\subsection{ Transmission properties of a Single Two-Level Atom Coupled to a Chiral Waveguide}
For a single two-level atom couple to a chiral waveguide, the system Hamiltonian is
\begin{equation}
	H=\Delta S^z+ik_{1}\alpha S^+-ik_1^*\alpha^*S^-,
\end{equation}
where we only consider the forward progation ($\varepsilon_{1\rightarrow}=\alpha$).

The quantum master equation for one two-level atom is
\begin{equation}
	\dot\rho=-i[H,\rho]-(\Gamma+\gamma)(S^+ S^-\rho-2 S^-\rho S^++\rho S^+ S^-),
\end{equation}
where $\Gamma=(|k_1|^2+|k_2|^2)/2$, and $\gamma$ is the intrinsic damping rate. The input-output relations are
\begin{equation}
	\begin{aligned}
	&\varepsilon_{1\leftarrow}=\varepsilon_{2\leftarrow}-k_{2}^*S^-,\\
	&\varepsilon_{2\rightarrow}=\varepsilon_{1\rightarrow}-k_{1}^*S^-.
	\end{aligned}
\end{equation}

For the steady state with $\dot\rho_{11}=\dot\rho_{12}=\dot\rho_{22}=0$, the field amplitude transmission and reflectivity are
\begin{equation}
	t=\frac{{\rm tr}(\varepsilon_{2\rightarrow}\rho)}{\alpha}=1-\frac{(-i\Delta+\Gamma+\gamma)|k_1|^2}{\Delta^2+(\Gamma+\gamma)^2+2|k_1|^2|\alpha|^2},
\end{equation}
\begin{equation}
	r=\frac{{\rm tr}(\varepsilon_{1\leftarrow}\rho)}{\alpha}=-\frac{(-i\Delta+\Gamma+\gamma)k_1k_2^*}{\Delta^2+(\Gamma+\gamma)^2+2|k_1|^2|\alpha|^2}.
\end{equation}
The power transmission and reflectivity are
\begin{equation}
		T=\frac{\langle \varepsilon_{2\rightarrow}^\dagger \varepsilon_{2\rightarrow}\rangle}{|\alpha|^2}=\frac{\Delta^2+\beta^2+2|k_1|^2|\alpha|^2}{\Delta^2+(\Gamma+\gamma)^2+2|k_1|^2|\alpha|^2},
\end{equation}
\begin{equation}
		R=\frac{\langle \varepsilon_{1\leftarrow}^\dagger \varepsilon_{1\leftarrow}\rangle}{|\alpha|^2}=\frac{|k_1|^2|k_2|^2}{\Delta^2+(\Gamma+\gamma)^2+2|k_1|^2|\alpha|^2},
\end{equation}
where we define $\beta=\gamma-\gamma_c$, with $\gamma_c=(|k_{1}|^2-|k_{2}|^2)/2$. The quantum fluctuation (incoherent part) of the transmitted field amplitude is
\begin{equation}
	T-|t|^2=\frac{2|k_1|^6|\alpha|^2}{[\Delta^2+(\Gamma+\gamma)^2+2|k_1|^2|\alpha|^2]^2}.
\end{equation}
For very small values of $|\alpha|$, the fluctuation scales as $T - |t|^2 \sim O(|\alpha|^2)$, consistent with the classical limit. The intensity of the field leaked out of the waveguide is given by
\begin{equation}
	1-T-R=\frac{2|k_1|^2\gamma}{\Delta^2+(\Gamma+\gamma)^2+2|k_1|^2|\alpha|^2},
\end{equation}
demonstrating its direct proportionality to the intrinsic damping $\gamma$.

\subsection{Quantum Critical Coupling Condition}
\begin{figure}[h]
	\centering
	\includegraphics[width=9cm]{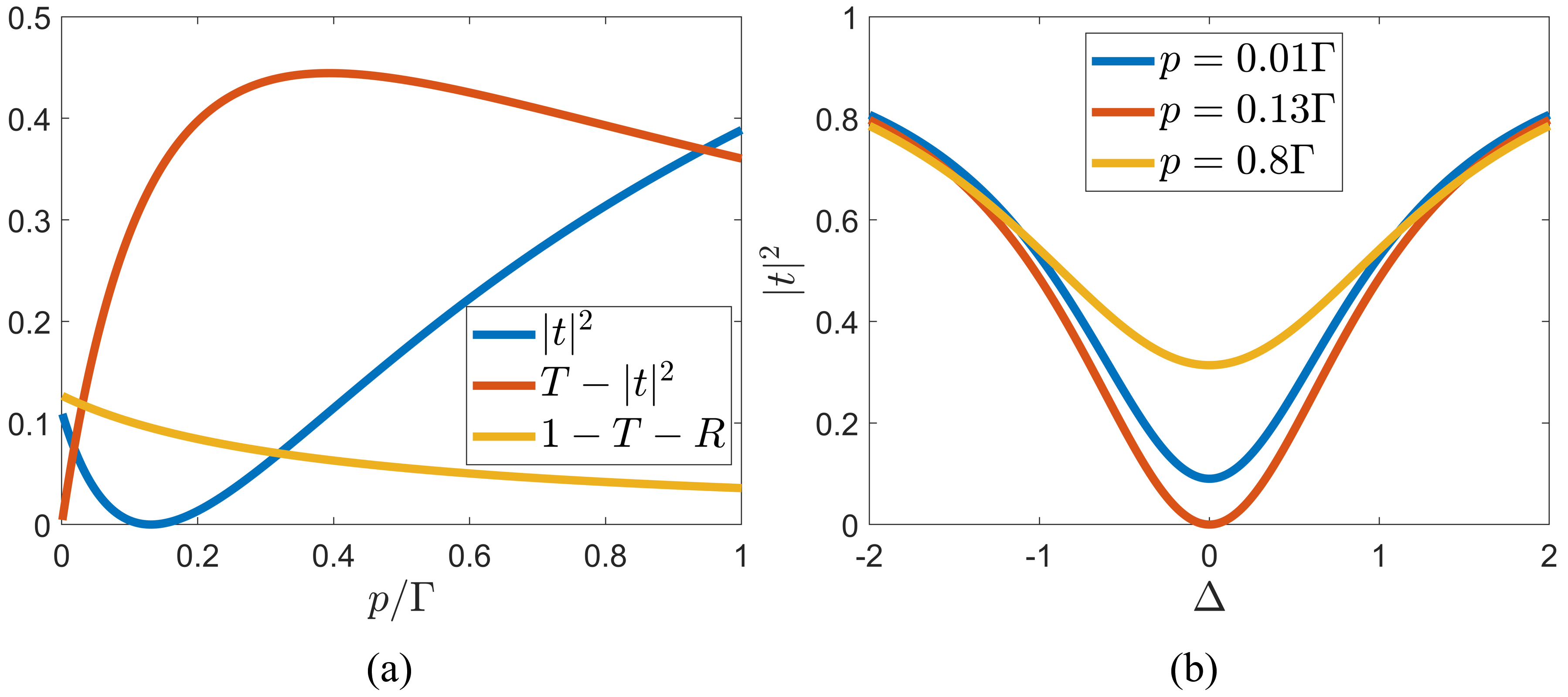} 
	\caption{(a) Field amplitude transmission $|t|^2$, quantum fluctuation $T-|t|^2$ and leakage $1-T-R$ as functions of input power $p$ for a single two-level atom. Parameters are $\Delta=0$, $\gamma=0.05\Gamma$, $k_1=\sqrt{1.4\Gamma}$, $k_2=\sqrt{0.6\Gamma}$. The dip in $|t|^2$ occurs at $p\simeq0.13$, consistent with the prediction from Eq. (\ref{qnc1}). (b) $|t|^2$ as a function of detuning $\Delta$ for various input power levels $p$. At $p = 0.13$ and $\Delta = 0$, the transmission $|t|^2$ its minimum value, approaching zero. For power levels deviating from the critical coupling condition, the transmission at $\Delta = 0$ increases, moving away from zero.}
	\label{srp}
\end{figure}
When considering one classical harmonic oscillator, the cavity is critically coupled when $\gamma-(|k_{1}|^2-|k_{2}|^2)/2=0$, then the field amplitude transmission and reflectivity are
\begin{equation}
	t=\frac{i\Delta}{i\Delta+|k_{1}|^2},
\end{equation}
\begin{equation}
	r=-\frac{k_{1}k_{2}^*}{i\Delta+|k_{1}|^2}.
\end{equation}
For a two-level atom,
\begin{equation}
	|t|^2=\frac{[\Delta^2+\beta(\Gamma+\gamma)+2|k_1|^2|\alpha|^2]^2+|k_1|^4\Delta^2}{[\Delta^2+(\Gamma+\gamma)^2+2|k_1|^2|\alpha|^2]^2}.
\end{equation}
Under the classical critical coupling condition ($\beta = 0$) and for $\Delta = 0$, the field amplitude transmission is
\begin{equation}
	|t|^2=\frac{4|k_1|^4|\alpha|^4}{[(\Gamma+\gamma)^2+2|k_1|^2|\alpha|^2]^2},
\end{equation}
and the power transmission is
\begin{equation}
	T=\frac{2|k_1|^2|\alpha|^2}{(\Gamma+\gamma)^2+2|k_1|^2|\alpha|^2}.
\end{equation}
For very small values of $|\alpha|$, $|t|^2 \sim O(|\alpha|^4)$, $T \sim O(|\alpha|^2)$, which recovers the classical limit. If $\beta<0$, $|\alpha|$ must have a specific value to ensure that $|t|=0$. That's the new feature of the quantum nonlinear system. For quantum nonlinear case, the critical coupling condition must be
\begin{equation}
	\beta(\Gamma+\gamma)+2|k_1|^2|\alpha|^2=0,\qquad \Delta=0,
\end{equation}
that is, the drive power should be
\begin{equation}
	p=\frac{(\frac{|k_1|^2+|k_2|^2}{2}+\gamma)(\frac{|k_1|^2-|k_2|^2}{2}-\gamma)}{2|k_1|^2},
	\label{qnc1}
\end{equation}
and critical coupling can happen only when
\begin{equation}
	\beta\le0,\qquad \Delta=0.
	\label{qnc2}
\end{equation}
For $|\alpha|>0$, $T>0$. When $|\alpha|>0$ and critical coupling happens, $\Delta=0$, $|\alpha|^2=-\frac{(\Gamma+\gamma)\beta}{2|k_1|^2}>0$, $\beta<0$ from Eqs. (\ref{qnc1}) and (\ref{qnc2}),
\begin{equation}
	\begin{aligned}
		T&=\frac{\langle \varepsilon_{2\rightarrow}^\dagger \varepsilon_{2\rightarrow}\rangle}{|\alpha|^2}=\frac{-\beta}{\Gamma+\gamma}>0.
	\end{aligned}
\end{equation}
Field amplitude transmission, quantum fluctuations, and leakage in a single two-level atom system are plotted against the input power in Fig. \ref{srp}(a). The dip in $|t|^2$, occurring at $p\simeq0.13$, marks the critical coupling condition where coherent transmission is minimized. This phenomenon can be directly calculated from Eq. (\ref{qnc1}). In Fig. \ref{srp}(b), the effect of driving power on transmission as a function of detuning $\Delta$ is plotted. At the critical power $p=0.13$, $|t|^2$ reaches its minimum at $\Delta=0$. As the input power deviates from this critical point, the minimum transmission at $\Delta=0$ increases, reflecting a departure from the critical coupling condition.

\subsection{Second-Order Correlation Function $g^{(2)}_T(0)$ for the Transmitted Field}
As discussed in Section \ref{antib}, a single two-level atom coupled to a chiral waveguide cannot emit two photons simultaneously in the reflected field, resulting in $\langle \varepsilon_{1\leftarrow}^\dagger \varepsilon_{1\leftarrow}^\dagger \varepsilon_{1\leftarrow} \varepsilon_{1\leftarrow} \rangle = 0$, regardless of the input power. In contrast, the second-order correlation function for the transmitted field, $g^{(2)}_T(0)$, is nonzero due to the direct contribution of the input coherent field mixed into the output. For the transmitted field,
\begin{equation}
	\begin{aligned}
		g^{(2)}_T(0)&=\frac{\langle\varepsilon_{2\rightarrow}^\dagger\varepsilon_{2\rightarrow}^\dagger\varepsilon_{2\rightarrow}\varepsilon_{2\rightarrow}\rangle}{\langle\varepsilon_{2\rightarrow}^\dagger\varepsilon_{2\rightarrow}\rangle^2}\\
		&\!\!\!\!\!\!\!\!\!\!\!\!\!\!\!\!\!\!=1+\frac{[2\Delta^2+|k_1|^4-2(\Gamma+\gamma-|k_1|^2)^2+4|k_1|^2|\alpha|^2]|k_1|^4}{[\Delta^2+2|k_1|^2\alpha^2+(\Gamma+\gamma-|k_1|^2)^2]^2}.
	\end{aligned}
\end{equation}
When $|\alpha|>0$ and critical coupling happens, $\Delta=0$, $|\alpha|^2=-\frac{(\Gamma+\gamma)\beta}{2|k_1|^2}>0$, $\beta<0$,
\begin{equation}
	g^{(2)}_T(0)=\frac{(|k_1|^2-3\beta)(\Gamma+\gamma)}{\beta^2}.
\end{equation}
$g^{(2)}_T(0)$ becomes significantly large as $\beta$ approaches $0$.

\bibliography{main}

\begin{thebibliography}{72}%
\makeatletter
\providecommand \@ifxundefined [1]{%
 \@ifx{#1\undefined}
}%
\providecommand \@ifnum [1]{%
 \ifnum #1\expandafter \@firstoftwo
 \else \expandafter \@secondoftwo
 \fi
}%
\providecommand \@ifx [1]{%
 \ifx #1\expandafter \@firstoftwo
 \else \expandafter \@secondoftwo
 \fi
}%
\providecommand \natexlab [1]{#1}%
\providecommand \enquote  [1]{``#1''}%
\providecommand \bibnamefont  [1]{#1}%
\providecommand \bibfnamefont [1]{#1}%
\providecommand \citenamefont [1]{#1}%
\providecommand \href@noop [0]{\@secondoftwo}%
\providecommand \href [0]{\begingroup \@sanitize@url \@href}%
\providecommand \@href[1]{\@@startlink{#1}\@@href}%
\providecommand \@@href[1]{\endgroup#1\@@endlink}%
\providecommand \@sanitize@url [0]{\catcode `\\12\catcode `\$12\catcode
  `\&12\catcode `\#12\catcode `\^12\catcode `\_12\catcode `\%12\relax}%
\providecommand \@@startlink[1]{}%
\providecommand \@@endlink[0]{}%
\providecommand \url  [0]{\begingroup\@sanitize@url \@url }%
\providecommand \@url [1]{\endgroup\@href {#1}{\urlprefix }}%
\providecommand \urlprefix  [0]{URL }%
\providecommand \Eprint [0]{\href }%
\providecommand \doibase [0]{https://doi.org/}%
\providecommand \selectlanguage [0]{\@gobble}%
\providecommand \bibinfo  [0]{\@secondoftwo}%
\providecommand \bibfield  [0]{\@secondoftwo}%
\providecommand \translation [1]{[#1]}%
\providecommand \BibitemOpen [0]{}%
\providecommand \bibitemStop [0]{}%
\providecommand \bibitemNoStop [0]{.\EOS\space}%
\providecommand \EOS [0]{\spacefactor3000\relax}%
\providecommand \BibitemShut  [1]{\csname bibitem#1\endcsname}%
\let\auto@bib@innerbib\@empty
\bibitem [{\citenamefont {Lodahl}\ \emph {et~al.}(2017)\citenamefont {Lodahl},
  \citenamefont {Mahmoodian}, \citenamefont {Stobbe}, \citenamefont
  {Rauschenbeutel}, \citenamefont {Schneeweiss}, \citenamefont {Volz},
  \citenamefont {Pichler},\ and\ \citenamefont {Zoller}}]{lodahl2017chiral}%
  \BibitemOpen
  \bibfield  {author} {\bibinfo {author} {\bibfnamefont {P.}~\bibnamefont
  {Lodahl}}, \bibinfo {author} {\bibfnamefont {S.}~\bibnamefont {Mahmoodian}},
  \bibinfo {author} {\bibfnamefont {S.}~\bibnamefont {Stobbe}}, \bibinfo
  {author} {\bibfnamefont {A.}~\bibnamefont {Rauschenbeutel}}, \bibinfo
  {author} {\bibfnamefont {P.}~\bibnamefont {Schneeweiss}}, \bibinfo {author}
  {\bibfnamefont {J.}~\bibnamefont {Volz}}, \bibinfo {author} {\bibfnamefont
  {H.}~\bibnamefont {Pichler}},\ and\ \bibinfo {author} {\bibfnamefont
  {P.}~\bibnamefont {Zoller}},\ }\bibfield  {title} {\bibinfo {title} {Chiral
  quantum optics},\ }\href@noop {} {\bibfield  {journal} {\bibinfo  {journal}
  {Nature}\ }\textbf {\bibinfo {volume} {541}},\ \bibinfo {pages} {473}
  (\bibinfo {year} {2017})}\BibitemShut {NoStop}%
\bibitem [{\citenamefont {Bliokh}\ and\ \citenamefont
  {Nori}(2012)}]{bliokh2012transverse}%
  \BibitemOpen
  \bibfield  {author} {\bibinfo {author} {\bibfnamefont {K.~Y.}\ \bibnamefont
  {Bliokh}}\ and\ \bibinfo {author} {\bibfnamefont {F.}~\bibnamefont {Nori}},\
  }\bibfield  {title} {\bibinfo {title} {Transverse spin of a surface
  polariton},\ }\href@noop {} {\bibfield  {journal} {\bibinfo  {journal} {Phys.
  Rev. A}\ }\textbf {\bibinfo {volume} {85}},\ \bibinfo {pages} {061801}
  (\bibinfo {year} {2012})}\BibitemShut {NoStop}%
\bibitem [{\citenamefont {Bliokh}\ \emph
  {et~al.}(2015{\natexlab{a}})\citenamefont {Bliokh}, \citenamefont
  {Rodr{\'\i}guez-Fortu{\~n}o}, \citenamefont {Nori},\ and\ \citenamefont
  {Zayats}}]{bliokh2015spin}%
  \BibitemOpen
  \bibfield  {author} {\bibinfo {author} {\bibfnamefont {K.~Y.}\ \bibnamefont
  {Bliokh}}, \bibinfo {author} {\bibfnamefont {F.~J.}\ \bibnamefont
  {Rodr{\'\i}guez-Fortu{\~n}o}}, \bibinfo {author} {\bibfnamefont
  {F.}~\bibnamefont {Nori}},\ and\ \bibinfo {author} {\bibfnamefont {A.~V.}\
  \bibnamefont {Zayats}},\ }\bibfield  {title} {\bibinfo {title} {Spin--orbit
  interactions of light},\ }\href@noop {} {\bibfield  {journal} {\bibinfo
  {journal} {Nat. Photonics}\ }\textbf {\bibinfo {volume} {9}},\ \bibinfo
  {pages} {796} (\bibinfo {year} {2015}{\natexlab{a}})}\BibitemShut {NoStop}%
\bibitem [{\citenamefont {Aiello}\ \emph {et~al.}(2015)\citenamefont {Aiello},
  \citenamefont {Banzer}, \citenamefont {Neugebauer},\ and\ \citenamefont
  {Leuchs}}]{aiello2015transverse}%
  \BibitemOpen
  \bibfield  {author} {\bibinfo {author} {\bibfnamefont {A.}~\bibnamefont
  {Aiello}}, \bibinfo {author} {\bibfnamefont {P.}~\bibnamefont {Banzer}},
  \bibinfo {author} {\bibfnamefont {M.}~\bibnamefont {Neugebauer}},\ and\
  \bibinfo {author} {\bibfnamefont {G.}~\bibnamefont {Leuchs}},\ }\bibfield
  {title} {\bibinfo {title} {From transverse angular momentum to photonic
  wheels},\ }\href@noop {} {\bibfield  {journal} {\bibinfo  {journal} {Nat.
  Photonics}\ }\textbf {\bibinfo {volume} {9}},\ \bibinfo {pages} {789}
  (\bibinfo {year} {2015})}\BibitemShut {NoStop}%
\bibitem [{\citenamefont {Bliokh}\ \emph
  {et~al.}(2015{\natexlab{b}})\citenamefont {Bliokh}, \citenamefont
  {Smirnova},\ and\ \citenamefont {Nori}}]{bliokh2015quantum}%
  \BibitemOpen
  \bibfield  {author} {\bibinfo {author} {\bibfnamefont {K.~Y.}\ \bibnamefont
  {Bliokh}}, \bibinfo {author} {\bibfnamefont {D.}~\bibnamefont {Smirnova}},\
  and\ \bibinfo {author} {\bibfnamefont {F.}~\bibnamefont {Nori}},\ }\bibfield
  {title} {\bibinfo {title} {Quantum spin hall effect of light},\ }\href@noop
  {} {\bibfield  {journal} {\bibinfo  {journal} {Science}\ }\textbf {\bibinfo
  {volume} {348}},\ \bibinfo {pages} {1448} (\bibinfo {year}
  {2015}{\natexlab{b}})}\BibitemShut {NoStop}%
\bibitem [{\citenamefont {Luxmoore}\ \emph
  {et~al.}(2013{\natexlab{a}})\citenamefont {Luxmoore}, \citenamefont {Wasley},
  \citenamefont {Ramsay}, \citenamefont {Thijssen}, \citenamefont {Oulton},
  \citenamefont {Hugues}, \citenamefont {Kasture}, \citenamefont {Achanta},
  \citenamefont {Fox},\ and\ \citenamefont
  {Skolnick}}]{luxmoore2013interfacing}%
  \BibitemOpen
  \bibfield  {author} {\bibinfo {author} {\bibfnamefont {I.}~\bibnamefont
  {Luxmoore}}, \bibinfo {author} {\bibfnamefont {N.}~\bibnamefont {Wasley}},
  \bibinfo {author} {\bibfnamefont {A.}~\bibnamefont {Ramsay}}, \bibinfo
  {author} {\bibfnamefont {A.}~\bibnamefont {Thijssen}}, \bibinfo {author}
  {\bibfnamefont {R.}~\bibnamefont {Oulton}}, \bibinfo {author} {\bibfnamefont
  {M.}~\bibnamefont {Hugues}}, \bibinfo {author} {\bibfnamefont
  {S.}~\bibnamefont {Kasture}}, \bibinfo {author} {\bibfnamefont
  {V.}~\bibnamefont {Achanta}}, \bibinfo {author} {\bibfnamefont
  {A.}~\bibnamefont {Fox}},\ and\ \bibinfo {author} {\bibfnamefont
  {M.}~\bibnamefont {Skolnick}},\ }\bibfield  {title} {\bibinfo {title}
  {Interfacing spins in an ingaas quantum dot to a semiconductor waveguide
  circuit using emitted photons},\ }\href@noop {} {\bibfield  {journal}
  {\bibinfo  {journal} {Phys. Rev. Lett.}\ }\textbf {\bibinfo {volume} {110}},\
  \bibinfo {pages} {037402} (\bibinfo {year} {2013}{\natexlab{a}})}\BibitemShut
  {NoStop}%
\bibitem [{\citenamefont {Junge}\ \emph {et~al.}(2013)\citenamefont {Junge},
  \citenamefont {O’shea}, \citenamefont {Volz},\ and\ \citenamefont
  {Rauschenbeutel}}]{junge2013strong}%
  \BibitemOpen
  \bibfield  {author} {\bibinfo {author} {\bibfnamefont {C.}~\bibnamefont
  {Junge}}, \bibinfo {author} {\bibfnamefont {D.}~\bibnamefont {O’shea}},
  \bibinfo {author} {\bibfnamefont {J.}~\bibnamefont {Volz}},\ and\ \bibinfo
  {author} {\bibfnamefont {A.}~\bibnamefont {Rauschenbeutel}},\ }\bibfield
  {title} {\bibinfo {title} {Strong coupling between single atoms and
  nontransversal photons},\ }\href@noop {} {\bibfield  {journal} {\bibinfo
  {journal} {Phys. Rev. Lett.}\ }\textbf {\bibinfo {volume} {110}},\ \bibinfo
  {pages} {213604} (\bibinfo {year} {2013})}\BibitemShut {NoStop}%
\bibitem [{\citenamefont {Luxmoore}\ \emph
  {et~al.}(2013{\natexlab{b}})\citenamefont {Luxmoore}, \citenamefont {Wasley},
  \citenamefont {Ramsay}, \citenamefont {Thijssen}, \citenamefont {Oulton},
  \citenamefont {Hugues}, \citenamefont {Fox},\ and\ \citenamefont
  {Skolnick}}]{luxmoore2013optical}%
  \BibitemOpen
  \bibfield  {author} {\bibinfo {author} {\bibfnamefont {I.}~\bibnamefont
  {Luxmoore}}, \bibinfo {author} {\bibfnamefont {N.~A.}\ \bibnamefont
  {Wasley}}, \bibinfo {author} {\bibfnamefont {A.~J.}\ \bibnamefont {Ramsay}},
  \bibinfo {author} {\bibfnamefont {A.}~\bibnamefont {Thijssen}}, \bibinfo
  {author} {\bibfnamefont {R.}~\bibnamefont {Oulton}}, \bibinfo {author}
  {\bibfnamefont {M.}~\bibnamefont {Hugues}}, \bibinfo {author} {\bibfnamefont
  {A.}~\bibnamefont {Fox}},\ and\ \bibinfo {author} {\bibfnamefont
  {M.}~\bibnamefont {Skolnick}},\ }\bibfield  {title} {\bibinfo {title}
  {Optical control of the emission direction of a quantum dot},\ }\href@noop {}
  {\bibfield  {journal} {\bibinfo  {journal} {Appl. Phys. Lett.}\ }\textbf
  {\bibinfo {volume} {103}} (\bibinfo {year} {2013}{\natexlab{b}})}\BibitemShut
  {NoStop}%
\bibitem [{\citenamefont {Neugebauer}\ \emph {et~al.}(2014)\citenamefont
  {Neugebauer}, \citenamefont {Bauer}, \citenamefont {Banzer},\ and\
  \citenamefont {Leuchs}}]{neugebauer2014polarization}%
  \BibitemOpen
  \bibfield  {author} {\bibinfo {author} {\bibfnamefont {M.}~\bibnamefont
  {Neugebauer}}, \bibinfo {author} {\bibfnamefont {T.}~\bibnamefont {Bauer}},
  \bibinfo {author} {\bibfnamefont {P.}~\bibnamefont {Banzer}},\ and\ \bibinfo
  {author} {\bibfnamefont {G.}~\bibnamefont {Leuchs}},\ }\bibfield  {title}
  {\bibinfo {title} {Polarization tailored light driven directional optical
  nanobeacon},\ }\href@noop {} {\bibfield  {journal} {\bibinfo  {journal} {Nano
  Lett.}\ }\textbf {\bibinfo {volume} {14}},\ \bibinfo {pages} {2546} (\bibinfo
  {year} {2014})}\BibitemShut {NoStop}%
\bibitem [{\citenamefont {Petersen}\ \emph {et~al.}(2014)\citenamefont
  {Petersen}, \citenamefont {Volz},\ and\ \citenamefont
  {Rauschenbeutel}}]{petersen2014chiral}%
  \BibitemOpen
  \bibfield  {author} {\bibinfo {author} {\bibfnamefont {J.}~\bibnamefont
  {Petersen}}, \bibinfo {author} {\bibfnamefont {J.}~\bibnamefont {Volz}},\
  and\ \bibinfo {author} {\bibfnamefont {A.}~\bibnamefont {Rauschenbeutel}},\
  }\bibfield  {title} {\bibinfo {title} {Chiral nanophotonic waveguide
  interface based on spin-orbit interaction of light},\ }\href@noop {}
  {\bibfield  {journal} {\bibinfo  {journal} {Science}\ }\textbf {\bibinfo
  {volume} {346}},\ \bibinfo {pages} {67} (\bibinfo {year} {2014})}\BibitemShut
  {NoStop}%
\bibitem [{\citenamefont {Rodr{\'\i}guez-Fortu{\~n}o}\ \emph
  {et~al.}(2014)\citenamefont {Rodr{\'\i}guez-Fortu{\~n}o}, \citenamefont
  {Barber-Sanz}, \citenamefont {Puerto}, \citenamefont {Griol},\ and\
  \citenamefont {Mart{\'\i}nez}}]{rodriguez2014resolving}%
  \BibitemOpen
  \bibfield  {author} {\bibinfo {author} {\bibfnamefont {F.~J.}\ \bibnamefont
  {Rodr{\'\i}guez-Fortu{\~n}o}}, \bibinfo {author} {\bibfnamefont
  {I.}~\bibnamefont {Barber-Sanz}}, \bibinfo {author} {\bibfnamefont
  {D.}~\bibnamefont {Puerto}}, \bibinfo {author} {\bibfnamefont
  {A.}~\bibnamefont {Griol}},\ and\ \bibinfo {author} {\bibfnamefont
  {A.}~\bibnamefont {Mart{\'\i}nez}},\ }\bibfield  {title} {\bibinfo {title}
  {Resolving light handedness with an on-chip silicon microdisk},\ }\href@noop
  {} {\bibfield  {journal} {\bibinfo  {journal} {ACS Photonics}\ }\textbf
  {\bibinfo {volume} {1}},\ \bibinfo {pages} {762} (\bibinfo {year}
  {2014})}\BibitemShut {NoStop}%
\bibitem [{\citenamefont {Shomroni}\ \emph {et~al.}(2014)\citenamefont
  {Shomroni}, \citenamefont {Rosenblum}, \citenamefont {Lovsky}, \citenamefont
  {Bechler}, \citenamefont {Guendelman},\ and\ \citenamefont
  {Dayan}}]{shomroni2014all}%
  \BibitemOpen
  \bibfield  {author} {\bibinfo {author} {\bibfnamefont {I.}~\bibnamefont
  {Shomroni}}, \bibinfo {author} {\bibfnamefont {S.}~\bibnamefont {Rosenblum}},
  \bibinfo {author} {\bibfnamefont {Y.}~\bibnamefont {Lovsky}}, \bibinfo
  {author} {\bibfnamefont {O.}~\bibnamefont {Bechler}}, \bibinfo {author}
  {\bibfnamefont {G.}~\bibnamefont {Guendelman}},\ and\ \bibinfo {author}
  {\bibfnamefont {B.}~\bibnamefont {Dayan}},\ }\bibfield  {title} {\bibinfo
  {title} {All-optical routing of single photons by a one-atom switch
  controlled by a single photon},\ }\href@noop {} {\bibfield  {journal}
  {\bibinfo  {journal} {Science}\ }\textbf {\bibinfo {volume} {345}},\ \bibinfo
  {pages} {903} (\bibinfo {year} {2014})}\BibitemShut {NoStop}%
\bibitem [{\citenamefont {S{\"o}llner}\ \emph {et~al.}(2015)\citenamefont
  {S{\"o}llner}, \citenamefont {Mahmoodian}, \citenamefont {Hansen},
  \citenamefont {Midolo}, \citenamefont {Javadi}, \citenamefont
  {Kir{\v{s}}ansk{\.e}}, \citenamefont {Pregnolato}, \citenamefont {El-Ella},
  \citenamefont {Lee}, \citenamefont {Song} \emph
  {et~al.}}]{sollner2015deterministic}%
  \BibitemOpen
  \bibfield  {author} {\bibinfo {author} {\bibfnamefont {I.}~\bibnamefont
  {S{\"o}llner}}, \bibinfo {author} {\bibfnamefont {S.}~\bibnamefont
  {Mahmoodian}}, \bibinfo {author} {\bibfnamefont {S.~L.}\ \bibnamefont
  {Hansen}}, \bibinfo {author} {\bibfnamefont {L.}~\bibnamefont {Midolo}},
  \bibinfo {author} {\bibfnamefont {A.}~\bibnamefont {Javadi}}, \bibinfo
  {author} {\bibfnamefont {G.}~\bibnamefont {Kir{\v{s}}ansk{\.e}}}, \bibinfo
  {author} {\bibfnamefont {T.}~\bibnamefont {Pregnolato}}, \bibinfo {author}
  {\bibfnamefont {H.}~\bibnamefont {El-Ella}}, \bibinfo {author} {\bibfnamefont
  {E.~H.}\ \bibnamefont {Lee}}, \bibinfo {author} {\bibfnamefont {J.~D.}\
  \bibnamefont {Song}}, \emph {et~al.},\ }\bibfield  {title} {\bibinfo {title}
  {Deterministic photon--emitter coupling in chiral photonic circuits},\
  }\href@noop {} {\bibfield  {journal} {\bibinfo  {journal} {Nat.
  Nanotechnol.}\ }\textbf {\bibinfo {volume} {10}},\ \bibinfo {pages} {775}
  (\bibinfo {year} {2015})}\BibitemShut {NoStop}%
\bibitem [{\citenamefont {Le~Feber}\ \emph {et~al.}(2015)\citenamefont
  {Le~Feber}, \citenamefont {Rotenberg},\ and\ \citenamefont
  {Kuipers}}]{le2015nanophotonic}%
  \BibitemOpen
  \bibfield  {author} {\bibinfo {author} {\bibfnamefont {B.}~\bibnamefont
  {Le~Feber}}, \bibinfo {author} {\bibfnamefont {N.}~\bibnamefont
  {Rotenberg}},\ and\ \bibinfo {author} {\bibfnamefont {L.}~\bibnamefont
  {Kuipers}},\ }\bibfield  {title} {\bibinfo {title} {Nanophotonic control of
  circular dipole emission},\ }\href@noop {} {\bibfield  {journal} {\bibinfo
  {journal} {Nat. Commun.}\ }\textbf {\bibinfo {volume} {6}},\ \bibinfo {pages}
  {6695} (\bibinfo {year} {2015})}\BibitemShut {NoStop}%
\bibitem [{\citenamefont {Coles}\ \emph {et~al.}(2016)\citenamefont {Coles},
  \citenamefont {Price}, \citenamefont {Dixon}, \citenamefont {Royall},
  \citenamefont {Clarke}, \citenamefont {Kok}, \citenamefont {Skolnick},
  \citenamefont {Fox},\ and\ \citenamefont {Makhonin}}]{coles2016chirality}%
  \BibitemOpen
  \bibfield  {author} {\bibinfo {author} {\bibfnamefont {R.}~\bibnamefont
  {Coles}}, \bibinfo {author} {\bibfnamefont {D.}~\bibnamefont {Price}},
  \bibinfo {author} {\bibfnamefont {J.}~\bibnamefont {Dixon}}, \bibinfo
  {author} {\bibfnamefont {B.}~\bibnamefont {Royall}}, \bibinfo {author}
  {\bibfnamefont {E.}~\bibnamefont {Clarke}}, \bibinfo {author} {\bibfnamefont
  {P.}~\bibnamefont {Kok}}, \bibinfo {author} {\bibfnamefont {M.}~\bibnamefont
  {Skolnick}}, \bibinfo {author} {\bibfnamefont {A.}~\bibnamefont {Fox}},\ and\
  \bibinfo {author} {\bibfnamefont {M.}~\bibnamefont {Makhonin}},\ }\bibfield
  {title} {\bibinfo {title} {Chirality of nanophotonic waveguide with embedded
  quantum emitter for unidirectional spin transfer},\ }\href@noop {} {\bibfield
   {journal} {\bibinfo  {journal} {Nat. Commun.}\ }\textbf {\bibinfo {volume}
  {7}},\ \bibinfo {pages} {11183} (\bibinfo {year} {2016})}\BibitemShut
  {NoStop}%
\bibitem [{\citenamefont {Lee}\ \emph {et~al.}(2012)\citenamefont {Lee},
  \citenamefont {Lee}, \citenamefont {Park}, \citenamefont {Oh}, \citenamefont
  {Lee}, \citenamefont {Kim},\ and\ \citenamefont {Lee}}]{lee2012role}%
  \BibitemOpen
  \bibfield  {author} {\bibinfo {author} {\bibfnamefont {S.-Y.}\ \bibnamefont
  {Lee}}, \bibinfo {author} {\bibfnamefont {I.-M.}\ \bibnamefont {Lee}},
  \bibinfo {author} {\bibfnamefont {J.}~\bibnamefont {Park}}, \bibinfo {author}
  {\bibfnamefont {S.}~\bibnamefont {Oh}}, \bibinfo {author} {\bibfnamefont
  {W.}~\bibnamefont {Lee}}, \bibinfo {author} {\bibfnamefont {K.-Y.}\
  \bibnamefont {Kim}},\ and\ \bibinfo {author} {\bibfnamefont {B.}~\bibnamefont
  {Lee}},\ }\bibfield  {title} {\bibinfo {title} {Role of magnetic induction
  currents in nanoslit excitation of surface plasmon polaritons},\ }\href@noop
  {} {\bibfield  {journal} {\bibinfo  {journal} {Phys. Rev. Lett.}\ }\textbf
  {\bibinfo {volume} {108}},\ \bibinfo {pages} {213907} (\bibinfo {year}
  {2012})}\BibitemShut {NoStop}%
\bibitem [{\citenamefont {Lin}\ \emph {et~al.}(2013)\citenamefont {Lin},
  \citenamefont {Mueller}, \citenamefont {Wang}, \citenamefont {Yuan},
  \citenamefont {Antoniou}, \citenamefont {Yuan},\ and\ \citenamefont
  {Capasso}}]{lin2013polarization}%
  \BibitemOpen
  \bibfield  {author} {\bibinfo {author} {\bibfnamefont {J.}~\bibnamefont
  {Lin}}, \bibinfo {author} {\bibfnamefont {J.~B.}\ \bibnamefont {Mueller}},
  \bibinfo {author} {\bibfnamefont {Q.}~\bibnamefont {Wang}}, \bibinfo {author}
  {\bibfnamefont {G.}~\bibnamefont {Yuan}}, \bibinfo {author} {\bibfnamefont
  {N.}~\bibnamefont {Antoniou}}, \bibinfo {author} {\bibfnamefont {X.-C.}\
  \bibnamefont {Yuan}},\ and\ \bibinfo {author} {\bibfnamefont
  {F.}~\bibnamefont {Capasso}},\ }\bibfield  {title} {\bibinfo {title}
  {Polarization-controlled tunable directional coupling of surface plasmon
  polaritons},\ }\href@noop {} {\bibfield  {journal} {\bibinfo  {journal}
  {Science}\ }\textbf {\bibinfo {volume} {340}},\ \bibinfo {pages} {331}
  (\bibinfo {year} {2013})}\BibitemShut {NoStop}%
\bibitem [{\citenamefont {Rodr{\'\i}guez-Fortu{\~n}o}\ \emph
  {et~al.}(2013)\citenamefont {Rodr{\'\i}guez-Fortu{\~n}o}, \citenamefont
  {Marino}, \citenamefont {Ginzburg}, \citenamefont {O’Connor}, \citenamefont
  {Mart{\'\i}nez}, \citenamefont {Wurtz},\ and\ \citenamefont
  {Zayats}}]{rodriguez2013near}%
  \BibitemOpen
  \bibfield  {author} {\bibinfo {author} {\bibfnamefont {F.~J.}\ \bibnamefont
  {Rodr{\'\i}guez-Fortu{\~n}o}}, \bibinfo {author} {\bibfnamefont
  {G.}~\bibnamefont {Marino}}, \bibinfo {author} {\bibfnamefont
  {P.}~\bibnamefont {Ginzburg}}, \bibinfo {author} {\bibfnamefont
  {D.}~\bibnamefont {O’Connor}}, \bibinfo {author} {\bibfnamefont
  {A.}~\bibnamefont {Mart{\'\i}nez}}, \bibinfo {author} {\bibfnamefont {G.~A.}\
  \bibnamefont {Wurtz}},\ and\ \bibinfo {author} {\bibfnamefont {A.~V.}\
  \bibnamefont {Zayats}},\ }\bibfield  {title} {\bibinfo {title} {Near-field
  interference for the unidirectional excitation of electromagnetic guided
  modes},\ }\href@noop {} {\bibfield  {journal} {\bibinfo  {journal} {Science}\
  }\textbf {\bibinfo {volume} {340}},\ \bibinfo {pages} {328} (\bibinfo {year}
  {2013})}\BibitemShut {NoStop}%
\bibitem [{\citenamefont {O’connor}\ \emph {et~al.}(2014)\citenamefont
  {O’connor}, \citenamefont {Ginzburg}, \citenamefont
  {Rodr{\'\i}guez-Fortu{\~n}o}, \citenamefont {Wurtz},\ and\ \citenamefont
  {Zayats}}]{o2014spin}%
  \BibitemOpen
  \bibfield  {author} {\bibinfo {author} {\bibfnamefont {D.}~\bibnamefont
  {O’connor}}, \bibinfo {author} {\bibfnamefont {P.}~\bibnamefont
  {Ginzburg}}, \bibinfo {author} {\bibfnamefont {F.~J.}\ \bibnamefont
  {Rodr{\'\i}guez-Fortu{\~n}o}}, \bibinfo {author} {\bibfnamefont {G.~A.}\
  \bibnamefont {Wurtz}},\ and\ \bibinfo {author} {\bibfnamefont {A.~V.}\
  \bibnamefont {Zayats}},\ }\bibfield  {title} {\bibinfo {title} {Spin--orbit
  coupling in surface plasmon scattering by nanostructures},\ }\href@noop {}
  {\bibfield  {journal} {\bibinfo  {journal} {Nat. Commun.}\ }\textbf {\bibinfo
  {volume} {5}},\ \bibinfo {pages} {5327} (\bibinfo {year} {2014})}\BibitemShut
  {NoStop}%
\bibitem [{\citenamefont {Roy}\ \emph {et~al.}(2017)\citenamefont {Roy},
  \citenamefont {Wilson},\ and\ \citenamefont
  {Firstenberg}}]{roy2017colloquium}%
  \BibitemOpen
  \bibfield  {author} {\bibinfo {author} {\bibfnamefont {D.}~\bibnamefont
  {Roy}}, \bibinfo {author} {\bibfnamefont {C.~M.}\ \bibnamefont {Wilson}},\
  and\ \bibinfo {author} {\bibfnamefont {O.}~\bibnamefont {Firstenberg}},\
  }\bibfield  {title} {\bibinfo {title} {Colloquium: Strongly interacting
  photons in one-dimensional continuum},\ }\href@noop {} {\bibfield  {journal}
  {\bibinfo  {journal} {Rev. Mod. Phys.}\ }\textbf {\bibinfo {volume} {89}},\
  \bibinfo {pages} {021001} (\bibinfo {year} {2017})}\BibitemShut {NoStop}%
\bibitem [{\citenamefont {Sheremet}\ \emph {et~al.}(2023)\citenamefont
  {Sheremet}, \citenamefont {Petrov}, \citenamefont {Iorsh}, \citenamefont
  {Poshakinskiy},\ and\ \citenamefont {Poddubny}}]{sheremet2023waveguide}%
  \BibitemOpen
  \bibfield  {author} {\bibinfo {author} {\bibfnamefont {A.~S.}\ \bibnamefont
  {Sheremet}}, \bibinfo {author} {\bibfnamefont {M.~I.}\ \bibnamefont
  {Petrov}}, \bibinfo {author} {\bibfnamefont {I.~V.}\ \bibnamefont {Iorsh}},
  \bibinfo {author} {\bibfnamefont {A.~V.}\ \bibnamefont {Poshakinskiy}},\ and\
  \bibinfo {author} {\bibfnamefont {A.~N.}\ \bibnamefont {Poddubny}},\
  }\bibfield  {title} {\bibinfo {title} {Waveguide quantum electrodynamics:
  Collective radiance and photon-photon correlations},\ }\href@noop {}
  {\bibfield  {journal} {\bibinfo  {journal} {Rev. Mod. Phys.}\ }\textbf
  {\bibinfo {volume} {95}},\ \bibinfo {pages} {015002} (\bibinfo {year}
  {2023})}\BibitemShut {NoStop}%
\bibitem [{\citenamefont {Kannan}\ \emph
  {et~al.}(2020{\natexlab{a}})\citenamefont {Kannan}, \citenamefont {Campbell},
  \citenamefont {Vasconcelos}, \citenamefont {Winik}, \citenamefont {Kim},
  \citenamefont {Kjaergaard}, \citenamefont {Krantz}, \citenamefont {Melville},
  \citenamefont {Niedzielski}, \citenamefont {Yoder} \emph
  {et~al.}}]{kannan2020generating}%
  \BibitemOpen
  \bibfield  {author} {\bibinfo {author} {\bibfnamefont {B.}~\bibnamefont
  {Kannan}}, \bibinfo {author} {\bibfnamefont {D.~L.}\ \bibnamefont
  {Campbell}}, \bibinfo {author} {\bibfnamefont {F.}~\bibnamefont
  {Vasconcelos}}, \bibinfo {author} {\bibfnamefont {R.}~\bibnamefont {Winik}},
  \bibinfo {author} {\bibfnamefont {D.}~\bibnamefont {Kim}}, \bibinfo {author}
  {\bibfnamefont {M.}~\bibnamefont {Kjaergaard}}, \bibinfo {author}
  {\bibfnamefont {P.}~\bibnamefont {Krantz}}, \bibinfo {author} {\bibfnamefont
  {A.}~\bibnamefont {Melville}}, \bibinfo {author} {\bibfnamefont {B.~M.}\
  \bibnamefont {Niedzielski}}, \bibinfo {author} {\bibfnamefont
  {J.}~\bibnamefont {Yoder}}, \emph {et~al.},\ }\bibfield  {title} {\bibinfo
  {title} {Generating spatially entangled itinerant photons with waveguide
  quantum electrodynamics},\ }\href@noop {} {\bibfield  {journal} {\bibinfo
  {journal} {Sci. Adv.}\ }\textbf {\bibinfo {volume} {6}},\ \bibinfo {pages}
  {eabb8780} (\bibinfo {year} {2020}{\natexlab{a}})}\BibitemShut {NoStop}%
\bibitem [{\citenamefont {Gonzalez-Ballestero}\ \emph
  {et~al.}(2014)\citenamefont {Gonzalez-Ballestero}, \citenamefont {Moreno},\
  and\ \citenamefont {Garcia-Vidal}}]{gonzalez2014generation}%
  \BibitemOpen
  \bibfield  {author} {\bibinfo {author} {\bibfnamefont {C.}~\bibnamefont
  {Gonzalez-Ballestero}}, \bibinfo {author} {\bibfnamefont {E.}~\bibnamefont
  {Moreno}},\ and\ \bibinfo {author} {\bibfnamefont {F.}~\bibnamefont
  {Garcia-Vidal}},\ }\bibfield  {title} {\bibinfo {title} {Generation,
  manipulation, and detection of two-qubit entanglement in waveguide qed},\
  }\href@noop {} {\bibfield  {journal} {\bibinfo  {journal} {Phys. Rev. A}\
  }\textbf {\bibinfo {volume} {89}},\ \bibinfo {pages} {042328} (\bibinfo
  {year} {2014})}\BibitemShut {NoStop}%
\bibitem [{\citenamefont {Facchi}\ \emph {et~al.}(2016)\citenamefont {Facchi},
  \citenamefont {Kim}, \citenamefont {Pascazio}, \citenamefont {Pepe},
  \citenamefont {Pomarico},\ and\ \citenamefont {Tufarelli}}]{facchi2016bound}%
  \BibitemOpen
  \bibfield  {author} {\bibinfo {author} {\bibfnamefont {P.}~\bibnamefont
  {Facchi}}, \bibinfo {author} {\bibfnamefont {M.}~\bibnamefont {Kim}},
  \bibinfo {author} {\bibfnamefont {S.}~\bibnamefont {Pascazio}}, \bibinfo
  {author} {\bibfnamefont {F.~V.}\ \bibnamefont {Pepe}}, \bibinfo {author}
  {\bibfnamefont {D.}~\bibnamefont {Pomarico}},\ and\ \bibinfo {author}
  {\bibfnamefont {T.}~\bibnamefont {Tufarelli}},\ }\bibfield  {title} {\bibinfo
  {title} {Bound states and entanglement generation in waveguide quantum
  electrodynamics},\ }\href@noop {} {\bibfield  {journal} {\bibinfo  {journal}
  {Phys. Rev. A}\ }\textbf {\bibinfo {volume} {94}},\ \bibinfo {pages} {043839}
  (\bibinfo {year} {2016})}\BibitemShut {NoStop}%
\bibitem [{\citenamefont {Shah}\ \emph {et~al.}(2024)\citenamefont {Shah},
  \citenamefont {Yang}, \citenamefont {Joshi},\ and\ \citenamefont
  {Mirhosseini}}]{shah2024stabilizing}%
  \BibitemOpen
  \bibfield  {author} {\bibinfo {author} {\bibfnamefont {P.~S.}\ \bibnamefont
  {Shah}}, \bibinfo {author} {\bibfnamefont {F.}~\bibnamefont {Yang}}, \bibinfo
  {author} {\bibfnamefont {C.}~\bibnamefont {Joshi}},\ and\ \bibinfo {author}
  {\bibfnamefont {M.}~\bibnamefont {Mirhosseini}},\ }\bibfield  {title}
  {\bibinfo {title} {Stabilizing remote entanglement via waveguide
  dissipation},\ }\href@noop {} {\bibfield  {journal} {\bibinfo  {journal} {PRX
  Quantum}\ }\textbf {\bibinfo {volume} {5}},\ \bibinfo {pages} {030346}
  (\bibinfo {year} {2024})}\BibitemShut {NoStop}%
\bibitem [{\citenamefont {Tiranov}\ \emph {et~al.}(2023)\citenamefont
  {Tiranov}, \citenamefont {Angelopoulou}, \citenamefont {van Diepen},
  \citenamefont {Schrinski}, \citenamefont {Sandberg}, \citenamefont {Wang},
  \citenamefont {Midolo}, \citenamefont {Scholz}, \citenamefont {Wieck},
  \citenamefont {Ludwig} \emph {et~al.}}]{tiranov2023collective}%
  \BibitemOpen
  \bibfield  {author} {\bibinfo {author} {\bibfnamefont {A.}~\bibnamefont
  {Tiranov}}, \bibinfo {author} {\bibfnamefont {V.}~\bibnamefont
  {Angelopoulou}}, \bibinfo {author} {\bibfnamefont {C.~J.}\ \bibnamefont {van
  Diepen}}, \bibinfo {author} {\bibfnamefont {B.}~\bibnamefont {Schrinski}},
  \bibinfo {author} {\bibfnamefont {O.~A.~D.}\ \bibnamefont {Sandberg}},
  \bibinfo {author} {\bibfnamefont {Y.}~\bibnamefont {Wang}}, \bibinfo {author}
  {\bibfnamefont {L.}~\bibnamefont {Midolo}}, \bibinfo {author} {\bibfnamefont
  {S.}~\bibnamefont {Scholz}}, \bibinfo {author} {\bibfnamefont {A.~D.}\
  \bibnamefont {Wieck}}, \bibinfo {author} {\bibfnamefont {A.}~\bibnamefont
  {Ludwig}}, \emph {et~al.},\ }\bibfield  {title} {\bibinfo {title} {Collective
  super-and subradiant dynamics between distant optical quantum emitters},\
  }\href@noop {} {\bibfield  {journal} {\bibinfo  {journal} {Science}\ }\textbf
  {\bibinfo {volume} {379}},\ \bibinfo {pages} {389} (\bibinfo {year}
  {2023})}\BibitemShut {NoStop}%
\bibitem [{\citenamefont {You}\ \emph {et~al.}(2018)\citenamefont {You},
  \citenamefont {Liao}, \citenamefont {Li},\ and\ \citenamefont
  {Zubairy}}]{you2018waveguide}%
  \BibitemOpen
  \bibfield  {author} {\bibinfo {author} {\bibfnamefont {J.}~\bibnamefont
  {You}}, \bibinfo {author} {\bibfnamefont {Z.}~\bibnamefont {Liao}}, \bibinfo
  {author} {\bibfnamefont {S.-W.}\ \bibnamefont {Li}},\ and\ \bibinfo {author}
  {\bibfnamefont {M.~S.}\ \bibnamefont {Zubairy}},\ }\bibfield  {title}
  {\bibinfo {title} {Waveguide quantum electrodynamics in squeezed vacuum},\
  }\href@noop {} {\bibfield  {journal} {\bibinfo  {journal} {Phys. Rev. A}\
  }\textbf {\bibinfo {volume} {97}},\ \bibinfo {pages} {023810} (\bibinfo
  {year} {2018})}\BibitemShut {NoStop}%
\bibitem [{\citenamefont {Goban}\ \emph {et~al.}(2015)\citenamefont {Goban},
  \citenamefont {Hung}, \citenamefont {Hood}, \citenamefont {Yu}, \citenamefont
  {Muniz}, \citenamefont {Painter},\ and\ \citenamefont
  {Kimble}}]{goban2015superradiance}%
  \BibitemOpen
  \bibfield  {author} {\bibinfo {author} {\bibfnamefont {A.}~\bibnamefont
  {Goban}}, \bibinfo {author} {\bibfnamefont {C.-L.}\ \bibnamefont {Hung}},
  \bibinfo {author} {\bibfnamefont {J.}~\bibnamefont {Hood}}, \bibinfo {author}
  {\bibfnamefont {S.-P.}\ \bibnamefont {Yu}}, \bibinfo {author} {\bibfnamefont
  {J.}~\bibnamefont {Muniz}}, \bibinfo {author} {\bibfnamefont
  {O.}~\bibnamefont {Painter}},\ and\ \bibinfo {author} {\bibfnamefont
  {H.}~\bibnamefont {Kimble}},\ }\bibfield  {title} {\bibinfo {title}
  {Superradiance for atoms trapped along a photonic crystal waveguide},\
  }\href@noop {} {\bibfield  {journal} {\bibinfo  {journal} {Phys. Rev. Lett.}\
  }\textbf {\bibinfo {volume} {115}},\ \bibinfo {pages} {063601} (\bibinfo
  {year} {2015})}\BibitemShut {NoStop}%
\bibitem [{\citenamefont {Shen}\ and\ \citenamefont
  {Fan}(2009)}]{shen2009theory}%
  \BibitemOpen
  \bibfield  {author} {\bibinfo {author} {\bibfnamefont {J.-T.}\ \bibnamefont
  {Shen}}\ and\ \bibinfo {author} {\bibfnamefont {S.}~\bibnamefont {Fan}},\
  }\bibfield  {title} {\bibinfo {title} {Theory of single-photon transport in a
  single-mode waveguide. i. coupling to a cavity containing a two-level atom},\
  }\href@noop {} {\bibfield  {journal} {\bibinfo  {journal} {Phys. Rev. A}\
  }\textbf {\bibinfo {volume} {79}},\ \bibinfo {pages} {023837} (\bibinfo
  {year} {2009})}\BibitemShut {NoStop}%
\bibitem [{\citenamefont {Liao}\ \emph {et~al.}(2015)\citenamefont {Liao},
  \citenamefont {Zeng}, \citenamefont {Zhu},\ and\ \citenamefont
  {Zubairy}}]{liao2015single}%
  \BibitemOpen
  \bibfield  {author} {\bibinfo {author} {\bibfnamefont {Z.}~\bibnamefont
  {Liao}}, \bibinfo {author} {\bibfnamefont {X.}~\bibnamefont {Zeng}}, \bibinfo
  {author} {\bibfnamefont {S.-Y.}\ \bibnamefont {Zhu}},\ and\ \bibinfo {author}
  {\bibfnamefont {M.~S.}\ \bibnamefont {Zubairy}},\ }\bibfield  {title}
  {\bibinfo {title} {Single-photon transport through an atomic chain coupled to
  a one-dimensional nanophotonic waveguide},\ }\href@noop {} {\bibfield
  {journal} {\bibinfo  {journal} {Phys. Rev. A}\ }\textbf {\bibinfo {volume}
  {92}},\ \bibinfo {pages} {023806} (\bibinfo {year} {2015})}\BibitemShut
  {NoStop}%
\bibitem [{\citenamefont {Tsoi}\ and\ \citenamefont
  {Law}(2008)}]{tsoi2008quantum}%
  \BibitemOpen
  \bibfield  {author} {\bibinfo {author} {\bibfnamefont {T.}~\bibnamefont
  {Tsoi}}\ and\ \bibinfo {author} {\bibfnamefont {C.}~\bibnamefont {Law}},\
  }\bibfield  {title} {\bibinfo {title} {Quantum interference effects of a
  single photon interacting with an atomic chain inside a one-dimensional
  waveguide},\ }\href@noop {} {\bibfield  {journal} {\bibinfo  {journal} {Phys.
  Rev. A}\ }\textbf {\bibinfo {volume} {78}},\ \bibinfo {pages} {063832}
  (\bibinfo {year} {2008})}\BibitemShut {NoStop}%
\bibitem [{\citenamefont {Derouault}\ and\ \citenamefont
  {Bouch{\`e}ne}(2014)}]{derouault2014one}%
  \BibitemOpen
  \bibfield  {author} {\bibinfo {author} {\bibfnamefont {S.}~\bibnamefont
  {Derouault}}\ and\ \bibinfo {author} {\bibfnamefont {M.~A.}\ \bibnamefont
  {Bouch{\`e}ne}},\ }\bibfield  {title} {\bibinfo {title} {One-photon wave
  packet interacting with two separated atoms in a one-dimensional waveguide:
  Influence of virtual photons},\ }\href@noop {} {\bibfield  {journal}
  {\bibinfo  {journal} {Phys. Rev. A}\ }\textbf {\bibinfo {volume} {90}},\
  \bibinfo {pages} {023828} (\bibinfo {year} {2014})}\BibitemShut {NoStop}%
\bibitem [{\citenamefont {Cheng}\ \emph {et~al.}(2017)\citenamefont {Cheng},
  \citenamefont {Xu},\ and\ \citenamefont {Agarwal}}]{cheng2017waveguide}%
  \BibitemOpen
  \bibfield  {author} {\bibinfo {author} {\bibfnamefont {M.-T.}\ \bibnamefont
  {Cheng}}, \bibinfo {author} {\bibfnamefont {J.}~\bibnamefont {Xu}},\ and\
  \bibinfo {author} {\bibfnamefont {G.~S.}\ \bibnamefont {Agarwal}},\
  }\bibfield  {title} {\bibinfo {title} {Waveguide transport mediated by strong
  coupling with atoms},\ }\href@noop {} {\bibfield  {journal} {\bibinfo
  {journal} {Phys. Rev. A}\ }\textbf {\bibinfo {volume} {95}},\ \bibinfo
  {pages} {053807} (\bibinfo {year} {2017})}\BibitemShut {NoStop}%
\bibitem [{\citenamefont {Konyk}\ and\ \citenamefont
  {Gea-Banacloche}(2017)}]{konyk2017one}%
  \BibitemOpen
  \bibfield  {author} {\bibinfo {author} {\bibfnamefont {W.}~\bibnamefont
  {Konyk}}\ and\ \bibinfo {author} {\bibfnamefont {J.}~\bibnamefont
  {Gea-Banacloche}},\ }\bibfield  {title} {\bibinfo {title} {One-and two-photon
  scattering by two atoms in a waveguide},\ }\href@noop {} {\bibfield
  {journal} {\bibinfo  {journal} {Phys. Rev. A}\ }\textbf {\bibinfo {volume}
  {96}},\ \bibinfo {pages} {063826} (\bibinfo {year} {2017})}\BibitemShut
  {NoStop}%
\bibitem [{\citenamefont {Le~Kien}\ and\ \citenamefont
  {Rauschenbeutel}(2014)}]{le2014propagation}%
  \BibitemOpen
  \bibfield  {author} {\bibinfo {author} {\bibfnamefont {F.}~\bibnamefont
  {Le~Kien}}\ and\ \bibinfo {author} {\bibfnamefont {A.}~\bibnamefont
  {Rauschenbeutel}},\ }\bibfield  {title} {\bibinfo {title} {Propagation of
  nanofiber-guided light through an array of atoms},\ }\href@noop {} {\bibfield
   {journal} {\bibinfo  {journal} {Phys. Rev. A}\ }\textbf {\bibinfo {volume}
  {90}},\ \bibinfo {pages} {063816} (\bibinfo {year} {2014})}\BibitemShut
  {NoStop}%
\bibitem [{\citenamefont {Huck}\ \emph {et~al.}(2011)\citenamefont {Huck},
  \citenamefont {Kumar}, \citenamefont {Shakoor},\ and\ \citenamefont
  {Andersen}}]{huck2011controlled}%
  \BibitemOpen
  \bibfield  {author} {\bibinfo {author} {\bibfnamefont {A.}~\bibnamefont
  {Huck}}, \bibinfo {author} {\bibfnamefont {S.}~\bibnamefont {Kumar}},
  \bibinfo {author} {\bibfnamefont {A.}~\bibnamefont {Shakoor}},\ and\ \bibinfo
  {author} {\bibfnamefont {U.~L.}\ \bibnamefont {Andersen}},\ }\bibfield
  {title} {\bibinfo {title} {Controlled coupling of a single nitrogen-vacancy
  center to a silver nanowire},\ }\href@noop {} {\bibfield  {journal} {\bibinfo
   {journal} {Phys. Rev. Lett.}\ }\textbf {\bibinfo {volume} {106}},\ \bibinfo
  {pages} {096801} (\bibinfo {year} {2011})}\BibitemShut {NoStop}%
\bibitem [{\citenamefont {Yalla}\ \emph {et~al.}(2014)\citenamefont {Yalla},
  \citenamefont {Sadgrove}, \citenamefont {Nayak},\ and\ \citenamefont
  {Hakuta}}]{yalla2014cavity}%
  \BibitemOpen
  \bibfield  {author} {\bibinfo {author} {\bibfnamefont {R.}~\bibnamefont
  {Yalla}}, \bibinfo {author} {\bibfnamefont {M.}~\bibnamefont {Sadgrove}},
  \bibinfo {author} {\bibfnamefont {K.~P.}\ \bibnamefont {Nayak}},\ and\
  \bibinfo {author} {\bibfnamefont {K.}~\bibnamefont {Hakuta}},\ }\bibfield
  {title} {\bibinfo {title} {Cavity quantum electrodynamics on a nanofiber
  using a composite photonic crystal cavity},\ }\href@noop {} {\bibfield
  {journal} {\bibinfo  {journal} {Phys. Rev. Lett.}\ }\textbf {\bibinfo
  {volume} {113}},\ \bibinfo {pages} {143601} (\bibinfo {year}
  {2014})}\BibitemShut {NoStop}%
\bibitem [{\citenamefont {Javadi}\ \emph {et~al.}(2015)\citenamefont {Javadi},
  \citenamefont {S{\"o}llner}, \citenamefont {Arcari}, \citenamefont {Hansen},
  \citenamefont {Midolo}, \citenamefont {Mahmoodian}, \citenamefont
  {Kir{\v{s}}ansk{\.e}}, \citenamefont {Pregnolato}, \citenamefont {Lee},
  \citenamefont {Song} \emph {et~al.}}]{javadi2015single}%
  \BibitemOpen
  \bibfield  {author} {\bibinfo {author} {\bibfnamefont {A.}~\bibnamefont
  {Javadi}}, \bibinfo {author} {\bibfnamefont {I.}~\bibnamefont {S{\"o}llner}},
  \bibinfo {author} {\bibfnamefont {M.}~\bibnamefont {Arcari}}, \bibinfo
  {author} {\bibfnamefont {S.~L.}\ \bibnamefont {Hansen}}, \bibinfo {author}
  {\bibfnamefont {L.}~\bibnamefont {Midolo}}, \bibinfo {author} {\bibfnamefont
  {S.}~\bibnamefont {Mahmoodian}}, \bibinfo {author} {\bibfnamefont
  {G.}~\bibnamefont {Kir{\v{s}}ansk{\.e}}}, \bibinfo {author} {\bibfnamefont
  {T.}~\bibnamefont {Pregnolato}}, \bibinfo {author} {\bibfnamefont
  {E.}~\bibnamefont {Lee}}, \bibinfo {author} {\bibfnamefont {J.}~\bibnamefont
  {Song}}, \emph {et~al.},\ }\bibfield  {title} {\bibinfo {title}
  {Single-photon non-linear optics with a quantum dot in a waveguide},\
  }\href@noop {} {\bibfield  {journal} {\bibinfo  {journal} {Nat. Commun.}\
  }\textbf {\bibinfo {volume} {6}},\ \bibinfo {pages} {8655} (\bibinfo {year}
  {2015})}\BibitemShut {NoStop}%
\bibitem [{\citenamefont {Li}\ and\ \citenamefont {Wei}(2016)}]{li2016probing}%
  \BibitemOpen
  \bibfield  {author} {\bibinfo {author} {\bibfnamefont {X.}~\bibnamefont
  {Li}}\ and\ \bibinfo {author} {\bibfnamefont {L.}~\bibnamefont {Wei}},\
  }\bibfield  {title} {\bibinfo {title} {Probing a single dipolar interaction
  between a pair of two-level quantum system by scatterings of single photons
  in an aside waveguide},\ }\href@noop {} {\bibfield  {journal} {\bibinfo
  {journal} {Opt. Commun.}\ }\textbf {\bibinfo {volume} {366}},\ \bibinfo
  {pages} {163} (\bibinfo {year} {2016})}\BibitemShut {NoStop}%
\bibitem [{\citenamefont {Mukhopadhyay}\ and\ \citenamefont
  {Agarwal}(2020)}]{mukhopadhyay2020transparency}%
  \BibitemOpen
  \bibfield  {author} {\bibinfo {author} {\bibfnamefont {D.}~\bibnamefont
  {Mukhopadhyay}}\ and\ \bibinfo {author} {\bibfnamefont {G.~S.}\ \bibnamefont
  {Agarwal}},\ }\bibfield  {title} {\bibinfo {title} {Transparency in a chain
  of disparate quantum emitters strongly coupled to a waveguide},\ }\href@noop
  {} {\bibfield  {journal} {\bibinfo  {journal} {Phys. Rev. A}\ }\textbf
  {\bibinfo {volume} {101}},\ \bibinfo {pages} {063814} (\bibinfo {year}
  {2020})}\BibitemShut {NoStop}%
\bibitem [{\citenamefont {Liao}\ \emph {et~al.}(2016)\citenamefont {Liao},
  \citenamefont {Nha},\ and\ \citenamefont {Zubairy}}]{liao2016dynamical}%
  \BibitemOpen
  \bibfield  {author} {\bibinfo {author} {\bibfnamefont {Z.}~\bibnamefont
  {Liao}}, \bibinfo {author} {\bibfnamefont {H.}~\bibnamefont {Nha}},\ and\
  \bibinfo {author} {\bibfnamefont {M.~S.}\ \bibnamefont {Zubairy}},\
  }\bibfield  {title} {\bibinfo {title} {Dynamical theory of single-photon
  transport in a one-dimensional waveguide coupled to identical and
  nonidentical emitters},\ }\href@noop {} {\bibfield  {journal} {\bibinfo
  {journal} {Phys. Rev. A}\ }\textbf {\bibinfo {volume} {94}},\ \bibinfo
  {pages} {053842} (\bibinfo {year} {2016})}\BibitemShut {NoStop}%
\bibitem [{\citenamefont {Miao}\ and\ \citenamefont
  {Agarwal}(2024)}]{miao2024kerr}%
  \BibitemOpen
  \bibfield  {author} {\bibinfo {author} {\bibfnamefont {Q.}~\bibnamefont
  {Miao}}\ and\ \bibinfo {author} {\bibfnamefont {G.}~\bibnamefont {Agarwal}},\
  }\bibfield  {title} {\bibinfo {title} {Kerr nonlinearity induced
  nonreciprocity in dissipatively coupled resonators},\ }\href@noop {}
  {\bibfield  {journal} {\bibinfo  {journal} {Phys. Rev. Research}\ }\textbf
  {\bibinfo {volume} {6}},\ \bibinfo {pages} {033020} (\bibinfo {year}
  {2024})}\BibitemShut {NoStop}%
\bibitem [{\citenamefont {Wang}\ \emph {et~al.}(2019)\citenamefont {Wang},
  \citenamefont {Rao}, \citenamefont {Yang}, \citenamefont {Xu}, \citenamefont
  {Gui}, \citenamefont {Yao}, \citenamefont {You},\ and\ \citenamefont
  {Hu}}]{wang2019nonreciprocity}%
  \BibitemOpen
  \bibfield  {author} {\bibinfo {author} {\bibfnamefont {Y.-P.}\ \bibnamefont
  {Wang}}, \bibinfo {author} {\bibfnamefont {J.}~\bibnamefont {Rao}}, \bibinfo
  {author} {\bibfnamefont {Y.}~\bibnamefont {Yang}}, \bibinfo {author}
  {\bibfnamefont {P.-C.}\ \bibnamefont {Xu}}, \bibinfo {author} {\bibfnamefont
  {Y.}~\bibnamefont {Gui}}, \bibinfo {author} {\bibfnamefont {B.}~\bibnamefont
  {Yao}}, \bibinfo {author} {\bibfnamefont {J.}~\bibnamefont {You}},\ and\
  \bibinfo {author} {\bibfnamefont {C.-M.}\ \bibnamefont {Hu}},\ }\bibfield
  {title} {\bibinfo {title} {Nonreciprocity and unidirectional invisibility in
  cavity magnonics},\ }\href@noop {} {\bibfield  {journal} {\bibinfo  {journal}
  {Phys. Rev. Lett.}\ }\textbf {\bibinfo {volume} {123}},\ \bibinfo {pages}
  {127202} (\bibinfo {year} {2019})}\BibitemShut {NoStop}%
\bibitem [{\citenamefont {Fang}\ and\ \citenamefont
  {Baranger}(2017)}]{fang2017multiple}%
  \BibitemOpen
  \bibfield  {author} {\bibinfo {author} {\bibfnamefont {Y.-L.~L.}\
  \bibnamefont {Fang}}\ and\ \bibinfo {author} {\bibfnamefont {H.~U.}\
  \bibnamefont {Baranger}},\ }\bibfield  {title} {\bibinfo {title} {Multiple
  emitters in a waveguide: nonreciprocity and correlated photons at perfect
  elastic transmission},\ }\href@noop {} {\bibfield  {journal} {\bibinfo
  {journal} {Phys. Rev. A}\ }\textbf {\bibinfo {volume} {96}},\ \bibinfo
  {pages} {013842} (\bibinfo {year} {2017})}\BibitemShut {NoStop}%
\bibitem [{\citenamefont {Rosario~Hamann}\ \emph {et~al.}(2018)\citenamefont
  {Rosario~Hamann}, \citenamefont {M{\"u}ller}, \citenamefont {Jerger},
  \citenamefont {Zanner}, \citenamefont {Combes}, \citenamefont {Pletyukhov},
  \citenamefont {Weides}, \citenamefont {Stace},\ and\ \citenamefont
  {Fedorov}}]{rosario2018nonreciprocity}%
  \BibitemOpen
  \bibfield  {author} {\bibinfo {author} {\bibfnamefont {A.}~\bibnamefont
  {Rosario~Hamann}}, \bibinfo {author} {\bibfnamefont {C.}~\bibnamefont
  {M{\"u}ller}}, \bibinfo {author} {\bibfnamefont {M.}~\bibnamefont {Jerger}},
  \bibinfo {author} {\bibfnamefont {M.}~\bibnamefont {Zanner}}, \bibinfo
  {author} {\bibfnamefont {J.}~\bibnamefont {Combes}}, \bibinfo {author}
  {\bibfnamefont {M.}~\bibnamefont {Pletyukhov}}, \bibinfo {author}
  {\bibfnamefont {M.}~\bibnamefont {Weides}}, \bibinfo {author} {\bibfnamefont
  {T.~M.}\ \bibnamefont {Stace}},\ and\ \bibinfo {author} {\bibfnamefont
  {A.}~\bibnamefont {Fedorov}},\ }\bibfield  {title} {\bibinfo {title}
  {Nonreciprocity realized with quantum nonlinearity},\ }\href@noop {}
  {\bibfield  {journal} {\bibinfo  {journal} {Phys. Rev. Lett.}\ }\textbf
  {\bibinfo {volume} {121}},\ \bibinfo {pages} {123601} (\bibinfo {year}
  {2018})}\BibitemShut {NoStop}%
\bibitem [{\citenamefont {Sounas}\ and\ \citenamefont
  {Al{\`u}}(2017)}]{sounas2017non}%
  \BibitemOpen
  \bibfield  {author} {\bibinfo {author} {\bibfnamefont {D.~L.}\ \bibnamefont
  {Sounas}}\ and\ \bibinfo {author} {\bibfnamefont {A.}~\bibnamefont
  {Al{\`u}}},\ }\bibfield  {title} {\bibinfo {title} {Non-reciprocal photonics
  based on time modulation},\ }\href@noop {} {\bibfield  {journal} {\bibinfo
  {journal} {Nat. Photonics}\ }\textbf {\bibinfo {volume} {11}},\ \bibinfo
  {pages} {774} (\bibinfo {year} {2017})}\BibitemShut {NoStop}%
\bibitem [{\citenamefont {Cotrufo}\ \emph {et~al.}(2021)\citenamefont
  {Cotrufo}, \citenamefont {Mann}, \citenamefont {Moussa},\ and\ \citenamefont
  {Al{\`u}}}]{cotrufo2021nonlinearity}%
  \BibitemOpen
  \bibfield  {author} {\bibinfo {author} {\bibfnamefont {M.}~\bibnamefont
  {Cotrufo}}, \bibinfo {author} {\bibfnamefont {S.~A.}\ \bibnamefont {Mann}},
  \bibinfo {author} {\bibfnamefont {H.}~\bibnamefont {Moussa}},\ and\ \bibinfo
  {author} {\bibfnamefont {A.}~\bibnamefont {Al{\`u}}},\ }\bibfield  {title}
  {\bibinfo {title} {Nonlinearity-induced nonreciprocity—part i},\
  }\href@noop {} {\bibfield  {journal} {\bibinfo  {journal} {IEEE Trans.
  Microwave Theory Tech.}\ }\textbf {\bibinfo {volume} {69}},\ \bibinfo {pages}
  {3569} (\bibinfo {year} {2021})}\BibitemShut {NoStop}%
\bibitem [{\citenamefont {Claudon}\ \emph {et~al.}(2010)\citenamefont
  {Claudon}, \citenamefont {Bleuse}, \citenamefont {Malik}, \citenamefont
  {Bazin}, \citenamefont {Jaffrennou}, \citenamefont {Gregersen}, \citenamefont
  {Sauvan}, \citenamefont {Lalanne},\ and\ \citenamefont
  {G{\'e}rard}}]{claudon2010highly}%
  \BibitemOpen
  \bibfield  {author} {\bibinfo {author} {\bibfnamefont {J.}~\bibnamefont
  {Claudon}}, \bibinfo {author} {\bibfnamefont {J.}~\bibnamefont {Bleuse}},
  \bibinfo {author} {\bibfnamefont {N.~S.}\ \bibnamefont {Malik}}, \bibinfo
  {author} {\bibfnamefont {M.}~\bibnamefont {Bazin}}, \bibinfo {author}
  {\bibfnamefont {P.}~\bibnamefont {Jaffrennou}}, \bibinfo {author}
  {\bibfnamefont {N.}~\bibnamefont {Gregersen}}, \bibinfo {author}
  {\bibfnamefont {C.}~\bibnamefont {Sauvan}}, \bibinfo {author} {\bibfnamefont
  {P.}~\bibnamefont {Lalanne}},\ and\ \bibinfo {author} {\bibfnamefont {J.-M.}\
  \bibnamefont {G{\'e}rard}},\ }\bibfield  {title} {\bibinfo {title} {A highly
  efficient single-photon source based on a quantum dot in a photonic
  nanowire},\ }\href@noop {} {\bibfield  {journal} {\bibinfo  {journal} {Nat.
  Photonics}\ }\textbf {\bibinfo {volume} {4}},\ \bibinfo {pages} {174}
  (\bibinfo {year} {2010})}\BibitemShut {NoStop}%
\bibitem [{\citenamefont {Zhang}\ and\ \citenamefont
  {Baranger}(2018)}]{zhang2018quantum}%
  \BibitemOpen
  \bibfield  {author} {\bibinfo {author} {\bibfnamefont {X.~H.}\ \bibnamefont
  {Zhang}}\ and\ \bibinfo {author} {\bibfnamefont {H.~U.}\ \bibnamefont
  {Baranger}},\ }\bibfield  {title} {\bibinfo {title} {Quantum interference and
  complex photon statistics in waveguide qed},\ }\href@noop {} {\bibfield
  {journal} {\bibinfo  {journal} {Phys. Rev. A}\ }\textbf {\bibinfo {volume}
  {97}},\ \bibinfo {pages} {023813} (\bibinfo {year} {2018})}\BibitemShut
  {NoStop}%
\bibitem [{\citenamefont {Corzo}\ \emph {et~al.}(2016)\citenamefont {Corzo},
  \citenamefont {Gouraud}, \citenamefont {Chandra}, \citenamefont {Goban},
  \citenamefont {Sheremet}, \citenamefont {Kupriyanov},\ and\ \citenamefont
  {Laurat}}]{corzo2016large}%
  \BibitemOpen
  \bibfield  {author} {\bibinfo {author} {\bibfnamefont {N.~V.}\ \bibnamefont
  {Corzo}}, \bibinfo {author} {\bibfnamefont {B.}~\bibnamefont {Gouraud}},
  \bibinfo {author} {\bibfnamefont {A.}~\bibnamefont {Chandra}}, \bibinfo
  {author} {\bibfnamefont {A.}~\bibnamefont {Goban}}, \bibinfo {author}
  {\bibfnamefont {A.~S.}\ \bibnamefont {Sheremet}}, \bibinfo {author}
  {\bibfnamefont {D.~V.}\ \bibnamefont {Kupriyanov}},\ and\ \bibinfo {author}
  {\bibfnamefont {J.}~\bibnamefont {Laurat}},\ }\bibfield  {title} {\bibinfo
  {title} {Large bragg reflection from one-dimensional chains of trapped atoms
  near a nanoscale waveguide},\ }\href@noop {} {\bibfield  {journal} {\bibinfo
  {journal} {Phys. Rev. Lett.}\ }\textbf {\bibinfo {volume} {117}},\ \bibinfo
  {pages} {133603} (\bibinfo {year} {2016})}\BibitemShut {NoStop}%
\bibitem [{\citenamefont {S{\o}rensen}\ \emph {et~al.}(2016)\citenamefont
  {S{\o}rensen}, \citenamefont {B{\'e}guin}, \citenamefont {Kluge},
  \citenamefont {Iakoupov}, \citenamefont {S{\o}rensen}, \citenamefont
  {M{\"u}ller}, \citenamefont {Polzik},\ and\ \citenamefont
  {Appel}}]{sorensen2016coherent}%
  \BibitemOpen
  \bibfield  {author} {\bibinfo {author} {\bibfnamefont {H.}~\bibnamefont
  {S{\o}rensen}}, \bibinfo {author} {\bibfnamefont {J.-B.}\ \bibnamefont
  {B{\'e}guin}}, \bibinfo {author} {\bibfnamefont {K.}~\bibnamefont {Kluge}},
  \bibinfo {author} {\bibfnamefont {I.}~\bibnamefont {Iakoupov}}, \bibinfo
  {author} {\bibfnamefont {A.}~\bibnamefont {S{\o}rensen}}, \bibinfo {author}
  {\bibfnamefont {J.}~\bibnamefont {M{\"u}ller}}, \bibinfo {author}
  {\bibfnamefont {E.}~\bibnamefont {Polzik}},\ and\ \bibinfo {author}
  {\bibfnamefont {J.}~\bibnamefont {Appel}},\ }\bibfield  {title} {\bibinfo
  {title} {Coherent backscattering of light off one-dimensional atomic
  strings},\ }\href@noop {} {\bibfield  {journal} {\bibinfo  {journal} {Phys.
  Rev. Lett.}\ }\textbf {\bibinfo {volume} {117}},\ \bibinfo {pages} {133604}
  (\bibinfo {year} {2016})}\BibitemShut {NoStop}%
\bibitem [{\citenamefont {Corzo}\ \emph {et~al.}(2019)\citenamefont {Corzo},
  \citenamefont {Raskop}, \citenamefont {Chandra}, \citenamefont {Sheremet},
  \citenamefont {Gouraud},\ and\ \citenamefont {Laurat}}]{corzo2019waveguide}%
  \BibitemOpen
  \bibfield  {author} {\bibinfo {author} {\bibfnamefont {N.~V.}\ \bibnamefont
  {Corzo}}, \bibinfo {author} {\bibfnamefont {J.}~\bibnamefont {Raskop}},
  \bibinfo {author} {\bibfnamefont {A.}~\bibnamefont {Chandra}}, \bibinfo
  {author} {\bibfnamefont {A.~S.}\ \bibnamefont {Sheremet}}, \bibinfo {author}
  {\bibfnamefont {B.}~\bibnamefont {Gouraud}},\ and\ \bibinfo {author}
  {\bibfnamefont {J.}~\bibnamefont {Laurat}},\ }\bibfield  {title} {\bibinfo
  {title} {Waveguide-coupled single collective excitation of atomic arrays},\
  }\href@noop {} {\bibfield  {journal} {\bibinfo  {journal} {Nature}\ }\textbf
  {\bibinfo {volume} {566}},\ \bibinfo {pages} {359} (\bibinfo {year}
  {2019})}\BibitemShut {NoStop}%
\bibitem [{\citenamefont {Mitsch}\ \emph {et~al.}(2014)\citenamefont {Mitsch},
  \citenamefont {Sayrin}, \citenamefont {Albrecht}, \citenamefont
  {Schneeweiss},\ and\ \citenamefont {Rauschenbeutel}}]{mitsch2014quantum}%
  \BibitemOpen
  \bibfield  {author} {\bibinfo {author} {\bibfnamefont {R.}~\bibnamefont
  {Mitsch}}, \bibinfo {author} {\bibfnamefont {C.}~\bibnamefont {Sayrin}},
  \bibinfo {author} {\bibfnamefont {B.}~\bibnamefont {Albrecht}}, \bibinfo
  {author} {\bibfnamefont {P.}~\bibnamefont {Schneeweiss}},\ and\ \bibinfo
  {author} {\bibfnamefont {A.}~\bibnamefont {Rauschenbeutel}},\ }\bibfield
  {title} {\bibinfo {title} {Quantum state-controlled directional spontaneous
  emission of photons into a nanophotonic waveguide},\ }\href@noop {}
  {\bibfield  {journal} {\bibinfo  {journal} {Nat. Commun.}\ }\textbf {\bibinfo
  {volume} {5}},\ \bibinfo {pages} {5713} (\bibinfo {year} {2014})}\BibitemShut
  {NoStop}%
\bibitem [{\citenamefont {Kornovan}\ \emph {et~al.}(2017)\citenamefont
  {Kornovan}, \citenamefont {Petrov},\ and\ \citenamefont
  {Iorsh}}]{kornovan2017transport}%
  \BibitemOpen
  \bibfield  {author} {\bibinfo {author} {\bibfnamefont {D.}~\bibnamefont
  {Kornovan}}, \bibinfo {author} {\bibfnamefont {M.}~\bibnamefont {Petrov}},\
  and\ \bibinfo {author} {\bibfnamefont {I.}~\bibnamefont {Iorsh}},\ }\bibfield
   {title} {\bibinfo {title} {Transport and collective radiance in a basic
  quantum chiral optical model},\ }\href@noop {} {\bibfield  {journal}
  {\bibinfo  {journal} {Phys. Rev. B}\ }\textbf {\bibinfo {volume} {96}},\
  \bibinfo {pages} {115162} (\bibinfo {year} {2017})}\BibitemShut {NoStop}%
\bibitem [{\citenamefont {Mirza}\ \emph {et~al.}(2017)\citenamefont {Mirza},
  \citenamefont {Hoskins},\ and\ \citenamefont
  {Schotland}}]{mirza2017chirality}%
  \BibitemOpen
  \bibfield  {author} {\bibinfo {author} {\bibfnamefont {I.~M.}\ \bibnamefont
  {Mirza}}, \bibinfo {author} {\bibfnamefont {J.~G.}\ \bibnamefont {Hoskins}},\
  and\ \bibinfo {author} {\bibfnamefont {J.~C.}\ \bibnamefont {Schotland}},\
  }\bibfield  {title} {\bibinfo {title} {Chirality, band structure, and
  localization in waveguide quantum electrodynamics},\ }\href@noop {}
  {\bibfield  {journal} {\bibinfo  {journal} {Phys. Rev. A}\ }\textbf {\bibinfo
  {volume} {96}},\ \bibinfo {pages} {053804} (\bibinfo {year}
  {2017})}\BibitemShut {NoStop}%
\bibitem [{\citenamefont {Mirza}\ and\ \citenamefont
  {Schotland}(2016{\natexlab{a}})}]{mirza2016multiqubit}%
  \BibitemOpen
  \bibfield  {author} {\bibinfo {author} {\bibfnamefont {I.~M.}\ \bibnamefont
  {Mirza}}\ and\ \bibinfo {author} {\bibfnamefont {J.~C.}\ \bibnamefont
  {Schotland}},\ }\bibfield  {title} {\bibinfo {title} {Multiqubit entanglement
  in bidirectional-chiral-waveguide qed},\ }\href@noop {} {\bibfield  {journal}
  {\bibinfo  {journal} {Phys. Rev. A}\ }\textbf {\bibinfo {volume} {94}},\
  \bibinfo {pages} {012302} (\bibinfo {year} {2016}{\natexlab{a}})}\BibitemShut
  {NoStop}%
\bibitem [{\citenamefont {Mirza}\ and\ \citenamefont
  {Schotland}(2016{\natexlab{b}})}]{mirza2016two}%
  \BibitemOpen
  \bibfield  {author} {\bibinfo {author} {\bibfnamefont {I.~M.}\ \bibnamefont
  {Mirza}}\ and\ \bibinfo {author} {\bibfnamefont {J.~C.}\ \bibnamefont
  {Schotland}},\ }\bibfield  {title} {\bibinfo {title} {Two-photon entanglement
  in multiqubit bidirectional-waveguide qed},\ }\href@noop {} {\bibfield
  {journal} {\bibinfo  {journal} {Phys. Rev. A}\ }\textbf {\bibinfo {volume}
  {94}},\ \bibinfo {pages} {012309} (\bibinfo {year}
  {2016}{\natexlab{b}})}\BibitemShut {NoStop}%
\bibitem [{\citenamefont {Stannigel}\ \emph {et~al.}(2012)\citenamefont
  {Stannigel}, \citenamefont {Rabl},\ and\ \citenamefont
  {Zoller}}]{stannigel2012driven}%
  \BibitemOpen
  \bibfield  {author} {\bibinfo {author} {\bibfnamefont {K.}~\bibnamefont
  {Stannigel}}, \bibinfo {author} {\bibfnamefont {P.}~\bibnamefont {Rabl}},\
  and\ \bibinfo {author} {\bibfnamefont {P.}~\bibnamefont {Zoller}},\
  }\bibfield  {title} {\bibinfo {title} {Driven-dissipative preparation of
  entangled states in cascaded quantum-optical networks},\ }\href@noop {}
  {\bibfield  {journal} {\bibinfo  {journal} {New J. Phys.}\ }\textbf {\bibinfo
  {volume} {14}},\ \bibinfo {pages} {063014} (\bibinfo {year}
  {2012})}\BibitemShut {NoStop}%
\bibitem [{\citenamefont {Gonzalez-Ballestero}\ \emph
  {et~al.}(2015)\citenamefont {Gonzalez-Ballestero}, \citenamefont
  {Gonzalez-Tudela}, \citenamefont {Garcia-Vidal},\ and\ \citenamefont
  {Moreno}}]{gonzalez2015chiral}%
  \BibitemOpen
  \bibfield  {author} {\bibinfo {author} {\bibfnamefont {C.}~\bibnamefont
  {Gonzalez-Ballestero}}, \bibinfo {author} {\bibfnamefont {A.}~\bibnamefont
  {Gonzalez-Tudela}}, \bibinfo {author} {\bibfnamefont {F.~J.}\ \bibnamefont
  {Garcia-Vidal}},\ and\ \bibinfo {author} {\bibfnamefont {E.}~\bibnamefont
  {Moreno}},\ }\bibfield  {title} {\bibinfo {title} {Chiral route to
  spontaneous entanglement generation},\ }\href@noop {} {\bibfield  {journal}
  {\bibinfo  {journal} {Phys. Rev. B}\ }\textbf {\bibinfo {volume} {92}},\
  \bibinfo {pages} {155304} (\bibinfo {year} {2015})}\BibitemShut {NoStop}%
\bibitem [{\citenamefont {Pichler}\ \emph {et~al.}(2015)\citenamefont
  {Pichler}, \citenamefont {Ramos}, \citenamefont {Daley},\ and\ \citenamefont
  {Zoller}}]{pichler2015quantum}%
  \BibitemOpen
  \bibfield  {author} {\bibinfo {author} {\bibfnamefont {H.}~\bibnamefont
  {Pichler}}, \bibinfo {author} {\bibfnamefont {T.}~\bibnamefont {Ramos}},
  \bibinfo {author} {\bibfnamefont {A.~J.}\ \bibnamefont {Daley}},\ and\
  \bibinfo {author} {\bibfnamefont {P.}~\bibnamefont {Zoller}},\ }\bibfield
  {title} {\bibinfo {title} {Quantum optics of chiral spin networks},\
  }\href@noop {} {\bibfield  {journal} {\bibinfo  {journal} {Phys. Rev. A}\
  }\textbf {\bibinfo {volume} {91}},\ \bibinfo {pages} {042116} (\bibinfo
  {year} {2015})}\BibitemShut {NoStop}%
\bibitem [{\citenamefont {Ramos}\ \emph {et~al.}(2014)\citenamefont {Ramos},
  \citenamefont {Pichler}, \citenamefont {Daley},\ and\ \citenamefont
  {Zoller}}]{ramos2014quantum}%
  \BibitemOpen
  \bibfield  {author} {\bibinfo {author} {\bibfnamefont {T.}~\bibnamefont
  {Ramos}}, \bibinfo {author} {\bibfnamefont {H.}~\bibnamefont {Pichler}},
  \bibinfo {author} {\bibfnamefont {A.~J.}\ \bibnamefont {Daley}},\ and\
  \bibinfo {author} {\bibfnamefont {P.}~\bibnamefont {Zoller}},\ }\bibfield
  {title} {\bibinfo {title} {Quantum spin dimers from chiral dissipation in
  cold-atom chains},\ }\href@noop {} {\bibfield  {journal} {\bibinfo  {journal}
  {Phys. Rev. Lett.}\ }\textbf {\bibinfo {volume} {113}},\ \bibinfo {pages}
  {237203} (\bibinfo {year} {2014})}\BibitemShut {NoStop}%
\bibitem [{\citenamefont {Kockum}\ \emph {et~al.}(2018)\citenamefont {Kockum},
  \citenamefont {Johansson},\ and\ \citenamefont
  {Nori}}]{kockum2018decoherence}%
  \BibitemOpen
  \bibfield  {author} {\bibinfo {author} {\bibfnamefont {A.~F.}\ \bibnamefont
  {Kockum}}, \bibinfo {author} {\bibfnamefont {G.}~\bibnamefont {Johansson}},\
  and\ \bibinfo {author} {\bibfnamefont {F.}~\bibnamefont {Nori}},\ }\bibfield
  {title} {\bibinfo {title} {Decoherence-free interaction between giant atoms
  in waveguide quantum electrodynamics},\ }\href@noop {} {\bibfield  {journal}
  {\bibinfo  {journal} {Phys. Rev. Lett.}\ }\textbf {\bibinfo {volume} {120}},\
  \bibinfo {pages} {140404} (\bibinfo {year} {2018})}\BibitemShut {NoStop}%
\bibitem [{\citenamefont {Kannan}\ \emph
  {et~al.}(2020{\natexlab{b}})\citenamefont {Kannan}, \citenamefont
  {Ruckriegel}, \citenamefont {Campbell}, \citenamefont {Frisk~Kockum},
  \citenamefont {Braum{\"u}ller}, \citenamefont {Kim}, \citenamefont
  {Kjaergaard}, \citenamefont {Krantz}, \citenamefont {Melville}, \citenamefont
  {Niedzielski} \emph {et~al.}}]{kannan2020waveguide}%
  \BibitemOpen
  \bibfield  {author} {\bibinfo {author} {\bibfnamefont {B.}~\bibnamefont
  {Kannan}}, \bibinfo {author} {\bibfnamefont {M.~J.}\ \bibnamefont
  {Ruckriegel}}, \bibinfo {author} {\bibfnamefont {D.~L.}\ \bibnamefont
  {Campbell}}, \bibinfo {author} {\bibfnamefont {A.}~\bibnamefont
  {Frisk~Kockum}}, \bibinfo {author} {\bibfnamefont {J.}~\bibnamefont
  {Braum{\"u}ller}}, \bibinfo {author} {\bibfnamefont {D.~K.}\ \bibnamefont
  {Kim}}, \bibinfo {author} {\bibfnamefont {M.}~\bibnamefont {Kjaergaard}},
  \bibinfo {author} {\bibfnamefont {P.}~\bibnamefont {Krantz}}, \bibinfo
  {author} {\bibfnamefont {A.}~\bibnamefont {Melville}}, \bibinfo {author}
  {\bibfnamefont {B.~M.}\ \bibnamefont {Niedzielski}}, \emph {et~al.},\
  }\bibfield  {title} {\bibinfo {title} {Waveguide quantum electrodynamics with
  superconducting artificial giant atoms},\ }\href@noop {} {\bibfield
  {journal} {\bibinfo  {journal} {Nature}\ }\textbf {\bibinfo {volume} {583}},\
  \bibinfo {pages} {775} (\bibinfo {year} {2020}{\natexlab{b}})}\BibitemShut
  {NoStop}%
\bibitem [{\citenamefont {Wang}\ \emph {et~al.}(2021)\citenamefont {Wang},
  \citenamefont {Liu}, \citenamefont {Kockum}, \citenamefont {Li},\ and\
  \citenamefont {Nori}}]{wang2021tunable}%
  \BibitemOpen
  \bibfield  {author} {\bibinfo {author} {\bibfnamefont {X.}~\bibnamefont
  {Wang}}, \bibinfo {author} {\bibfnamefont {T.}~\bibnamefont {Liu}}, \bibinfo
  {author} {\bibfnamefont {A.~F.}\ \bibnamefont {Kockum}}, \bibinfo {author}
  {\bibfnamefont {H.-R.}\ \bibnamefont {Li}},\ and\ \bibinfo {author}
  {\bibfnamefont {F.}~\bibnamefont {Nori}},\ }\bibfield  {title} {\bibinfo
  {title} {Tunable chiral bound states with giant atoms},\ }\href@noop {}
  {\bibfield  {journal} {\bibinfo  {journal} {Phys. Rev. Lett.}\ }\textbf
  {\bibinfo {volume} {126}},\ \bibinfo {pages} {043602} (\bibinfo {year}
  {2021})}\BibitemShut {NoStop}%
\bibitem [{\citenamefont {Wang}\ \emph {et~al.}(2024)\citenamefont {Wang},
  \citenamefont {Zhu}, \citenamefont {Liu},\ and\ \citenamefont
  {Nori}}]{wang2024realizing}%
  \BibitemOpen
  \bibfield  {author} {\bibinfo {author} {\bibfnamefont {X.}~\bibnamefont
  {Wang}}, \bibinfo {author} {\bibfnamefont {H.-B.}\ \bibnamefont {Zhu}},
  \bibinfo {author} {\bibfnamefont {T.}~\bibnamefont {Liu}},\ and\ \bibinfo
  {author} {\bibfnamefont {F.}~\bibnamefont {Nori}},\ }\bibfield  {title}
  {\bibinfo {title} {Realizing quantum optics in structured environments with
  giant atoms},\ }\href@noop {} {\bibfield  {journal} {\bibinfo  {journal}
  {Phys. Rev. Research}\ }\textbf {\bibinfo {volume} {6}},\ \bibinfo {pages}
  {013279} (\bibinfo {year} {2024})}\BibitemShut {NoStop}%
\bibitem [{\citenamefont {Dong}\ \emph {et~al.}(2015)\citenamefont {Dong},
  \citenamefont {Shen}, \citenamefont {Zou}, \citenamefont {Zhang},
  \citenamefont {Fu},\ and\ \citenamefont {Guo}}]{dong2015brillouin}%
  \BibitemOpen
  \bibfield  {author} {\bibinfo {author} {\bibfnamefont {C.-H.}\ \bibnamefont
  {Dong}}, \bibinfo {author} {\bibfnamefont {Z.}~\bibnamefont {Shen}}, \bibinfo
  {author} {\bibfnamefont {C.-L.}\ \bibnamefont {Zou}}, \bibinfo {author}
  {\bibfnamefont {Y.-L.}\ \bibnamefont {Zhang}}, \bibinfo {author}
  {\bibfnamefont {W.}~\bibnamefont {Fu}},\ and\ \bibinfo {author}
  {\bibfnamefont {G.-C.}\ \bibnamefont {Guo}},\ }\bibfield  {title} {\bibinfo
  {title} {Brillouin-scattering-induced transparency and non-reciprocal light
  storage},\ }\href@noop {} {\bibfield  {journal} {\bibinfo  {journal} {Nat.
  Commun.}\ }\textbf {\bibinfo {volume} {6}},\ \bibinfo {pages} {6193}
  (\bibinfo {year} {2015})}\BibitemShut {NoStop}%
\bibitem [{\citenamefont {Kim}\ \emph {et~al.}(2015)\citenamefont {Kim},
  \citenamefont {Kuzyk}, \citenamefont {Han}, \citenamefont {Wang},\ and\
  \citenamefont {Bahl}}]{kim2015non}%
  \BibitemOpen
  \bibfield  {author} {\bibinfo {author} {\bibfnamefont {J.}~\bibnamefont
  {Kim}}, \bibinfo {author} {\bibfnamefont {M.~C.}\ \bibnamefont {Kuzyk}},
  \bibinfo {author} {\bibfnamefont {K.}~\bibnamefont {Han}}, \bibinfo {author}
  {\bibfnamefont {H.}~\bibnamefont {Wang}},\ and\ \bibinfo {author}
  {\bibfnamefont {G.}~\bibnamefont {Bahl}},\ }\bibfield  {title} {\bibinfo
  {title} {Non-reciprocal brillouin scattering induced transparency},\
  }\href@noop {} {\bibfield  {journal} {\bibinfo  {journal} {Nat. Phys.}\
  }\textbf {\bibinfo {volume} {11}},\ \bibinfo {pages} {275} (\bibinfo {year}
  {2015})}\BibitemShut {NoStop}%
\bibitem [{\citenamefont {Peterson}\ \emph {et~al.}(2019)\citenamefont
  {Peterson}, \citenamefont {Benalcazar}, \citenamefont {Lin}, \citenamefont
  {Hughes},\ and\ \citenamefont {Bahl}}]{peterson2019strong}%
  \BibitemOpen
  \bibfield  {author} {\bibinfo {author} {\bibfnamefont {C.~W.}\ \bibnamefont
  {Peterson}}, \bibinfo {author} {\bibfnamefont {W.~A.}\ \bibnamefont
  {Benalcazar}}, \bibinfo {author} {\bibfnamefont {M.}~\bibnamefont {Lin}},
  \bibinfo {author} {\bibfnamefont {T.~L.}\ \bibnamefont {Hughes}},\ and\
  \bibinfo {author} {\bibfnamefont {G.}~\bibnamefont {Bahl}},\ }\bibfield
  {title} {\bibinfo {title} {Strong nonreciprocity in modulated resonator
  chains through synthetic electric and magnetic fields},\ }\href@noop {}
  {\bibfield  {journal} {\bibinfo  {journal} {Phys. Rev. Lett.}\ }\textbf
  {\bibinfo {volume} {123}},\ \bibinfo {pages} {063901} (\bibinfo {year}
  {2019})}\BibitemShut {NoStop}%
\bibitem [{\citenamefont {Biehs}\ and\ \citenamefont
  {Agarwal}(2023)}]{biehs2023enhancement}%
  \BibitemOpen
  \bibfield  {author} {\bibinfo {author} {\bibfnamefont {S.-A.}\ \bibnamefont
  {Biehs}}\ and\ \bibinfo {author} {\bibfnamefont {G.}~\bibnamefont
  {Agarwal}},\ }\bibfield  {title} {\bibinfo {title} {Enhancement of synthetic
  magnetic field induced nonreciprocity via bound states in the continuum in
  dissipatively coupled systems},\ }\href@noop {} {\bibfield  {journal}
  {\bibinfo  {journal} {Phys. Rev. B}\ }\textbf {\bibinfo {volume} {108}},\
  \bibinfo {pages} {035423} (\bibinfo {year} {2023})}\BibitemShut {NoStop}%
\bibitem [{\citenamefont {Suh}\ \emph {et~al.}(2004)\citenamefont {Suh},
  \citenamefont {Wang},\ and\ \citenamefont {Fan}}]{suh2004temporal}%
  \BibitemOpen
  \bibfield  {author} {\bibinfo {author} {\bibfnamefont {W.}~\bibnamefont
  {Suh}}, \bibinfo {author} {\bibfnamefont {Z.}~\bibnamefont {Wang}},\ and\
  \bibinfo {author} {\bibfnamefont {S.}~\bibnamefont {Fan}},\ }\bibfield
  {title} {\bibinfo {title} {Temporal coupled-mode theory and the presence of
  non-orthogonal modes in lossless multimode cavities},\ }\href@noop {}
  {\bibfield  {journal} {\bibinfo  {journal} {IEEE J. Quantum Electron.}\
  }\textbf {\bibinfo {volume} {40}},\ \bibinfo {pages} {1511} (\bibinfo {year}
  {2004})}\BibitemShut {NoStop}%
\bibitem [{\citenamefont {Zhao}\ \emph {et~al.}(2019)\citenamefont {Zhao},
  \citenamefont {Guo},\ and\ \citenamefont {Fan}}]{zhao2019connection}%
  \BibitemOpen
  \bibfield  {author} {\bibinfo {author} {\bibfnamefont {Z.}~\bibnamefont
  {Zhao}}, \bibinfo {author} {\bibfnamefont {C.}~\bibnamefont {Guo}},\ and\
  \bibinfo {author} {\bibfnamefont {S.}~\bibnamefont {Fan}},\ }\bibfield
  {title} {\bibinfo {title} {Connection of temporal coupled-mode-theory
  formalisms for a resonant optical system and its time-reversal conjugate},\
  }\href@noop {} {\bibfield  {journal} {\bibinfo  {journal} {Phys. Rev. A}\
  }\textbf {\bibinfo {volume} {99}},\ \bibinfo {pages} {033839} (\bibinfo
  {year} {2019})}\BibitemShut {NoStop}%
\bibitem [{\citenamefont {Agarwal}(2012)}]{agarwal2012quantum}%
  \BibitemOpen
  \bibfield  {author} {\bibinfo {author} {\bibfnamefont {G.~S.}\ \bibnamefont
  {Agarwal}},\ }\href@noop {} {\emph {\bibinfo {title} {Quantum optics}}}\
  (\bibinfo  {publisher} {Cambridge University Press},\ \bibinfo {year}
  {2012})\BibitemShut {NoStop}%
\end{thebibliography}%

\end{document}